\documentclass[lettersize,journal]{IEEEtran}
\usepackage{amsmath,amsfonts}
\usepackage{array}
\usepackage[caption=false,font=normalsize,labelfont=sf,textfont=sf]{subfig}
\usepackage{textcomp}
\usepackage{stfloats}
\usepackage{url}
\usepackage{verbatim}
\usepackage{graphicx}
\usepackage{cite}
\usepackage[normalem]{ulem}
\usepackage{mathtools}
\usepackage{multirow}
\usepackage{tabularx}
\usepackage{subcaption}
\usepackage{enumitem}
\usepackage{tikz}
\newcommand{\ballnumber}[1]{\tikz[baseline=(myanchor.base)] \node[circle,fill=.,inner sep=1pt] (myanchor) {\color{-.}\bfseries\footnotesize #1};}

\usepackage[linesnumbered,ruled,vlined]{algorithm2e}

\newcommand{\ccblue}{\color{black}}

\begin{document}

\title{NL-DPE: An Analog In-memory Non-Linear Dot Product Engine for Efficient CNN and LLM Inference}

\author{Lei~Zhao,
        Luca~Buonanno,
        Archit~Gajjar,~\IEEEmembership{Member,~IEEE,},
        John~Moon,
        Aishwarya~Natarajan,~\IEEEmembership{Member,~IEEE,},
        Sergey~Serebryakov,
        Ron~M.~Roth,~\IEEEmembership{Life Fellow,~IEEE,},
        Xia~Sheng,
        Youtao~Zhang,~\IEEEmembership{Member,~IEEE,},
        Paolo~Faraboschi,~\IEEEmembership{Fellow,~IEEE,},
        Jim~Ignowski,~\IEEEmembership{Senior Member,~IEEE,},
        and~Giacomo~Pedretti,~\IEEEmembership{Member,~IEEE,}
\thanks{Lei~Zhao, Luca~Buonanno, Archit~Gajjar, John~Moon, Aishwarya~Natarajan, Sergey~Serebryakov, Xia~Sheng, Paolo~Faraboschi, Jim~Ignowski, and Giacomo~Pedretti are with the Artificial Intelligence Research Lab (AIRL), Hewlett
  Packard Labs, USA (e-mail: lei.zhao@hpe.com, giacomo.pedretti@hpe.com).}%
\thanks{Ronny~Roth is with Technion - Israel Institute of Technology.}%
\thanks{Youtao~Zhang is with the University of Pittsburgh, Pittsburgh, PA, USA.}%
}



\maketitle

\begin{abstract}
Resistive Random Access Memory (RRAM) based in-memory computing (IMC) accelerators offer significant performance and energy advantages for deep neural networks (DNNs), but face three major limitations:
(1) they support only \textit{static} dot-product operations and cannot accelerate arbitrary non-linear functions or data-dependent multiplications essential to modern LLMs;
(2) they demand large, power-hungry analog-to-digital converter (ADC) circuits;
and (3) mapping model weights to device conductance introduces errors from cell nonidealities.
These challenges hinder scalable and accurate IMC acceleration as models grow.

We propose NL-DPE, a Non-Linear Dot Product Engine that overcomes these barriers.
NL-DPE augments crosspoint arrays with RRAM-based Analog Content Addressable Memory (ACAM) to execute arbitrary non-linear functions and data-dependent matrix multiplications in the analog domain by transforming them into decision trees, fully eliminating ADCs.
To address device noise, NL-DPE uses software-based Noise Aware Fine-tuning (NAF), requiring no in-device calibration.
Experiments show that NL-DPE delivers 28$\times$ energy efficiency and 249$\times$ speedup over a GPU baseline, and 22$\times$ energy efficiency and 245$\times$ speedup over existing IMC accelerators, while maintaining high accuracy.
\end{abstract}

\begin{IEEEkeywords}
in-memory computing, analog acceleration, neural network, noise, fine-tuning, rram.
\end{IEEEkeywords}

\section{Introduction}

Artificial Intelligence (AI) models have significantly grown in size and complexity, 
leading to frequent data movement between memory and processing units.
The development of in-memory computing (IMC) accelerators, particularly those based on resistive memories, such as RRAM~\cite{wan2022compute}, Phase Change Memories (PCM)~\cite{le202364}, and FeFET~\cite{jerry2017ferroelectric}, stand out as one of the most promising solutions due to their potential for high energy efficiency and scalability.
IMC accelerators often organize resistive memory cells in a crossbar to construct dot product engines (DPEs), which accelerate vector-matrix multiplications (VMMs) in the analog domain, achieving low energy consumption and high parallelism \cite{shafiee2016isaac, ankit2019puma}.

However, modern AI models, in particular, the Transformer models that are widely adopted in large language models (LLMs), consist of more than just static VMM operations.
That is, they often require data-dependent matrix multiplications (DMMul) and non-linear operations, such as activations, and Softmax.
Most IMC accelerators~\cite{shafiee2016isaac,ankit2019puma,andrulis2023raella} perform activations and DMMul using on-chip digital circuits or the host processor, which demand the conversion of the analog outputs from DPEs into digital signals via analog-to-digital converters (ADCs) before computation.
Unfortunately, ADCs are both energy- and area-inefficient, consuming more than 30\% of the chip area and accounting for over 50\% of the total power consumption \cite{shafiee2016isaac, ankit2019puma}. 
Recently proposed IMC designs~\cite{song2021brahms,zhu2022fuse} strive to eliminate ADCs but work only if the AI model contains ReLU-like non-linear functions, and thus lack the flexibility to adapt to evolving AI models.

\begin{figure}[ht]
\begin{center}
\includegraphics[width=3in]{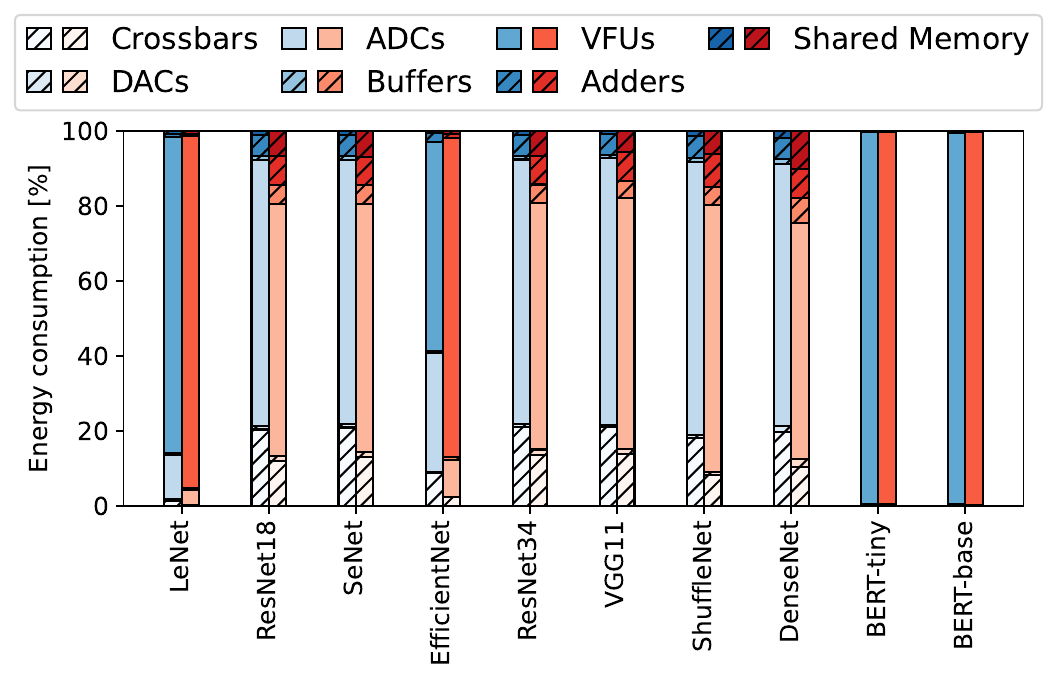}
\caption{ The energy breakdown of RRAM IMC accelerators, namely ISAAC~\cite{shafiee2016isaac} (blue) and RAELLA~\cite{andrulis2023raella} (red). 
We focus on optimizing ADC and VFU, i.e., the bars with solid colors.}
\vspace{-0.2in}
\label{fig:breakdown}
\end{center}
\end{figure}

{\ccblue
To understand the inefficiency of RRAM-based IMC accelerators for evolving AI models, we analyze the energy consumption breakdown of ISAAC~\cite{shafiee2016isaac} and RAELLA~\cite{andrulis2023raella}.}
Section \ref{sec:methodology} elaborates on the experiment settings.
For non-linear activation and DMMul operations, we optimize using a state-of-the-art Vector Functional Units (VFU)~\cite{reggiani2023flex,ankit2019puma} and simulate using CiMLoop~\cite{andrulis2024cimloop} \footnote{Alternatively, we may optimize to implement DMMul by frequently reprogramming the DPEs \cite{yang2020retransformer,yazdanbakhsh2022sparse}. Given the limited endurance of RRAM cells (e.g., $<10^9$ cycles~\cite{mannocci2023memory}), this approach is suboptimal and less preferred.}. 
Fig.~\ref{fig:breakdown} shows that while ADCs still account for the majority of the energy consumption for CNN models, VFUs dominate the energy consumption in the attention models, i.e., BERT~\cite{devlin2019bert}.
In summary, it is important to address the non-linear functions and DMMul in modern models.

Another challenge for IMC-based AI accelerators is that resistive memory cells, such as RRAMs, suffer from {\underline{conductance noises}}, i.e., there is a non-negligible difference between the desired cell conductance and the cell conductance that was actually programmed or can be read. 
While studies have shown that AI models exhibit inherent tolerance to various errors, including conductance noises, real hardware~\cite{sheng2019low,mao2022experimentally,wan2022compute} demonstrate that resistance noises play a crucial role in determining the model's accuracy. 
Without careful handling of various noise sources, model accuracy may degrade to unacceptable levels~\cite{mao2022experimentally}.
Existing approaches for mitigating conductance noises~\cite{wan2022compute} are limited to VMM operations and thus not usable for non-VMMs in the analog domain.

In this work, we propose NL-DPE, a noise-resilient non-linear dot product engine, to achieve scalable acceleration for modern AI models.
In particular, we design a RRAM-based Analog Content Address Memory (ACAM) to act as a decision-tree (DT) engine, enabling the acceleration of arbitrary non-linear functions in the analog domain.
Our contributions are as follows.

\begin{itemize}[noitemsep,nolistsep,leftmargin=*]
\item We enable the acceleration of both VMM and non-VMM operations in the analog domain.
In particular, DMMul and non-linear function operations are conducted in the ACAM engines.
By completely eliminating the costly ADC circuits, NL-DPE strives to achieve scalable performance and energy consumption improvements for modern AI models.

\item We propose a novel RRAM-based ACAM design that enables the computation of arbitrary non-linear functions in the analog domain, by predicting them with DTs.
In particular, the modular per-bit DT architecture provides sufficient accuracy to meet the diverse demands of modern AI models.
We elaborate on the decomposition of complicated functions, e.g., Softmax and DMMuls, into a sequence of logarithmic and exponential operations, which are then efficiently implemented in ACAM.

\item 
We achieve error-resilient deployment of large AI models using a software-based, per-DT Noise Aware Fine-tuning (NAF) approach.
From a pre-trained model, we extract DMMul and non-linear functions, convert them into DTs, and fine-tune each DT independently with simulated RRAM noise.
The resulting model can be directly deployed to NL-DPE arrays without any post-deployment calibration.

We measure noise from fabricated RRAM devices~\cite{sheng2019low}\footnote{Although the evaluation is based on Ta-Ox RRAM devices, the designs apply to other emerging resistive memory technologies, such as PCM and FeFETs.}, and develop noise models for both crossbar arrays and ACAM units that incorporate various RRAM non-idealities to facilitate NAF.

\item 
We evaluate NL-DPE using the CiMLoop simulator \cite{andrulis2024cimloop} and an NVIDIA H100 GPU.
NL-DPE delivers 28$\times$ higher energy efficiency and 249$\times$ speedup over the GPU baseline at the same quantization bit width, and 22$\times$ higher energy efficiency and 245$\times$ speedup compared to existing IMC accelerators.
A detailed design-space exploration further highlights the advantages of our architecture.

\end{itemize}

\section{Background}

\subsection{In-Memory Dot-Product Engine (DPE)} \label{sec:dpe}

\begin{figure}[t]
\begin{center}
\includegraphics[width=0.8\linewidth]{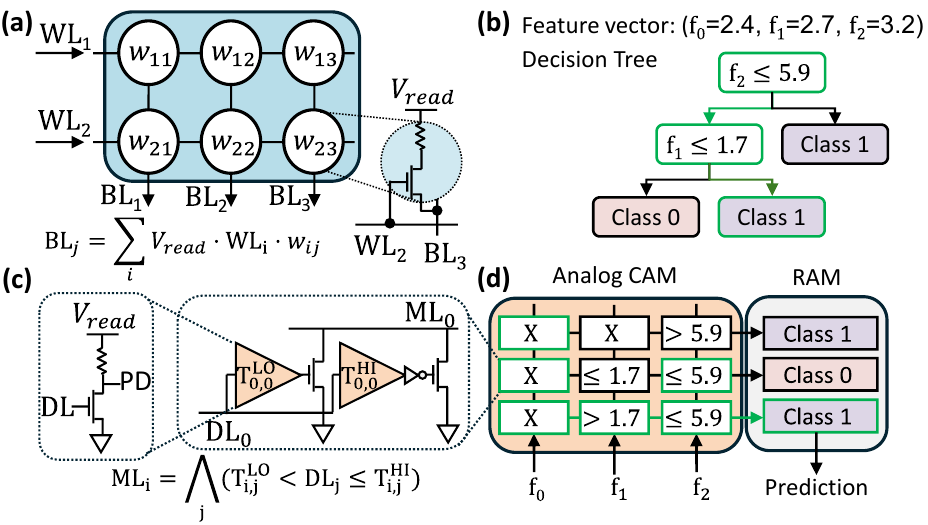}
\caption{\textbf{(a)} Computing the VMM of in a crossbar array with 1T1R RRAM cells. \textbf{(b)} Example of a trained decision tree. \textbf{(c)} ACAM cell using 1T1R \textbf{(d)} Mapping the decision tree of (a) onto ACAM and performing inference.}
\vspace{-0.2in}
\label{fig:dpe_acam}
\end{center}
\end{figure}

Fig.~\ref{fig:dpe_acam}(a) illustrates a VMM operation $\mathbf{y}=\mathbf{x}\mathbf{W}$ mapped onto a 2$\times$3 RRAM crossbar.
RRAM cells, typically in a 1T1R structure, store weights $w_{ij}$ as conductance values, while inputs $x_i$ are converted into sliced binary voltages on the word lines~\cite{shafiee2016isaac,andrulis2023raella}.
By Kirchhoff’s and Ohm’s laws, the bit-line currents represent the dot products scaled by the read voltage $V_{read}$, with column-wise accumulation completed via shift-and-add~\cite{shafiee2016isaac}.
This enables a full analog VMM in one step.
While the figure shows a small 2$\times$3 example, practical designs often use much larger arrays (e.g., 512×512~\cite{andrulis2023raella}) for high parallelism.

The analog outputs of the crossbar must be digitized for downstream operations such as activation functions, typically requiring area- and power-hungry ADCs.
Although pulse-width modulation can replace ADCs~\cite{jiang202240nm}, it introduces high latency due to unary encoding and still relies on digital activations.
A recent work also proposed a crossbar-based ramp ADC for recurrent neural networks~\cite{yang2025efficient}.

\subsection{Decision Tree}

A decision tree (DT) is a simple yet effective model for classification and regression and can serve as a universal function approximator.
It consists of internal nodes that apply conditional branches on trainable feature values and leaf nodes that produce the final prediction (e.g., a class label).

Fig.~\ref{fig:dpe_acam}(b) shows a DT trained for binary classification.
Each sample has three features $(f_0, f_1, f_2)$ and a label 0 or 1.
Internal nodes compare a selected feature to a threshold, and leaf nodes store the predicted label.

To classify an input, traversal starts at the root, following internal-node predicates to the left or right subtree until reaching a leaf containing the predicted label.
In Fig.~\ref{fig:dpe_acam}(b), the highlighted path shows the input $(f_0=2.4, f_1=2.7, f_2=3.2)$ reaching a leaf that predicts label 1.

\subsection{Analog Content Addressable Memory}

RRAM and FeFET have been used to implement ACAMs for accelerating tree-based ML algorithms in the analog domain~\cite{pedretti2021tree}.
Unlike digital CAMs, which compare single bits, an ACAM cell compares an analog input against a stored range.
In Fig.~\ref{fig:dpe_acam}(c), each cell uses two RRAMs to store a lower bound, upper bound, or a range, programmed as conductances ($\text{T}^{lo}{ij}$ and $\text{T}^{hi}{ij}$).
A wildcard cell ``X'' stores the maximum range for a always-TRUE comparison.
Analog comparators use a 1T1R voltage-divider structure, similar to crossbars, driving pull-down logic.
ACAM designs~\cite{li2020analog} store the lower bound with 1T1R and require an inverter for the upper bound.

Before searching, the match line ($\text{ML}$) is precharged high.
If the input $\text{DL}j$ is below $T^{lo}_{ij}$ or above $T^{hi}_{ij}$, the corresponding pull-down or inverter drives $\text{ML}_i$ low.
Thus, $\text{ML}_i$ stays high only when $\text{DL}_j$ falls within the stored range.

To map a DT onto an ACAM array, each leaf node is traced back to the root, storing feature thresholds along the path in a row of ACAM cells.
In Fig.~\ref{fig:dpe_acam}(d), the last row corresponds to the highlighted path in Fig.~\ref{fig:dpe_acam}(b).
Repeated features along a path are merged into a single range, and wildcard cells (“X”) cover features not on the path.
During inference, a row’s match line is activated if all input features fall within the stored ranges, retrieving the predicted class from adjacent RAM.
ACAMs thus accept analog inputs but produce digital match outputs.

\subsection{Attention Mechanism}

\begin{figure}[tbp]
\centerline{\includegraphics[width=3.4in]{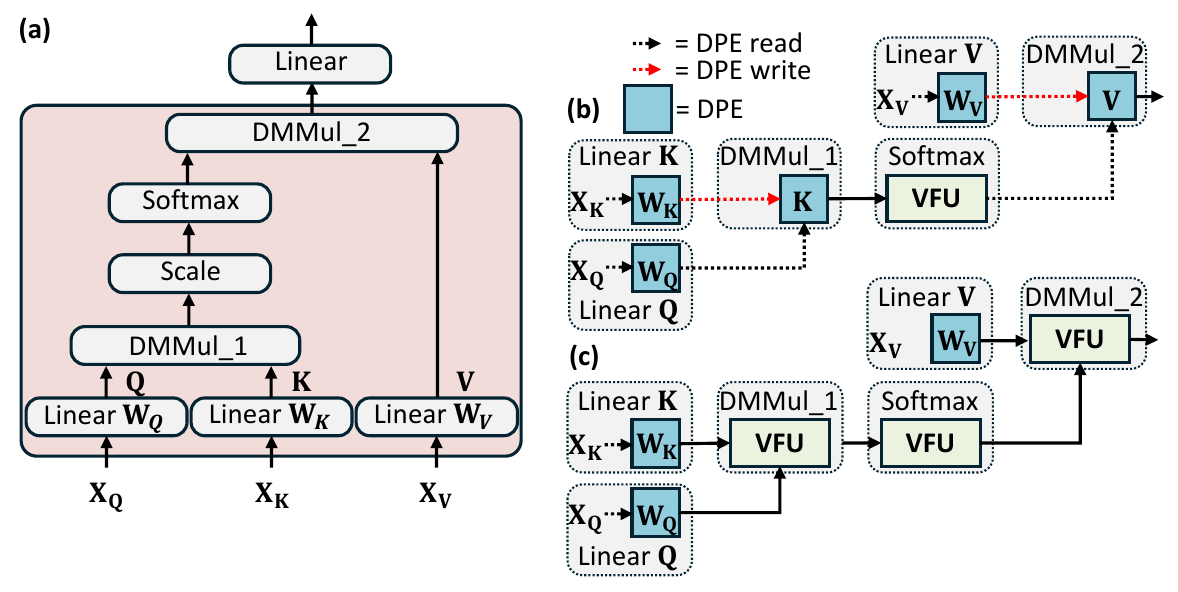}}
\caption{\textbf{(a)} Attention mechanism and its mapping into conventional DPE architectures either by \textbf{(b)} reprogramming the DPE arrays or \textbf{(c)} using a VFU for DMMul.}
\vspace{-0.1in}
\label{fig:attention}
\end{figure}

The attention mechanism is the core operation of Transformer models.
As shown in Fig.~\ref{fig:attention}, it takes three token sequences—query ($\mathbf{X_Q}$), key ($\mathbf{X_K}$), and value ($\mathbf{X_V}$))—which are each transformed by linear layers ($\mathbf{W_Q}$, $\mathbf{W_K}$, $\mathbf{W_V}$) into matrices $\mathbf{Q}$, $\mathbf{K}$, and $\mathbf{V}$.
Attention is then computed as:
\begin{equation}
Attention(\mathbf{K},\mathbf{Q},\mathbf{V}) = \text{Softmax}\left(\frac{\mathbf{Q} \cdot \mathbf{K}^T}{\sqrt{d_k}}\right) \cdot \mathbf{V}
\label{eq:attention}
\end{equation}
where $d_k$ represents the dimensionality of each query and key vector.
The Softmax function is defined as:
\begin{equation}
\text{Softmax}(y_i)=\frac{e^{y_i}}{\sum_{_j=1}^{L}e^{y_j}}
\label{eq:softmax}
\end{equation}
Here, $y_i$ is an element in the input vector $\mathbf{Y} = [y_1, ..., y_L]$ for Softmax, and $L$ denotes the vector length.
The output of the attention mechanism is then passed to another Linear layer.

\begin{figure*}[ht]
\begin{center}
\includegraphics[width=7in]{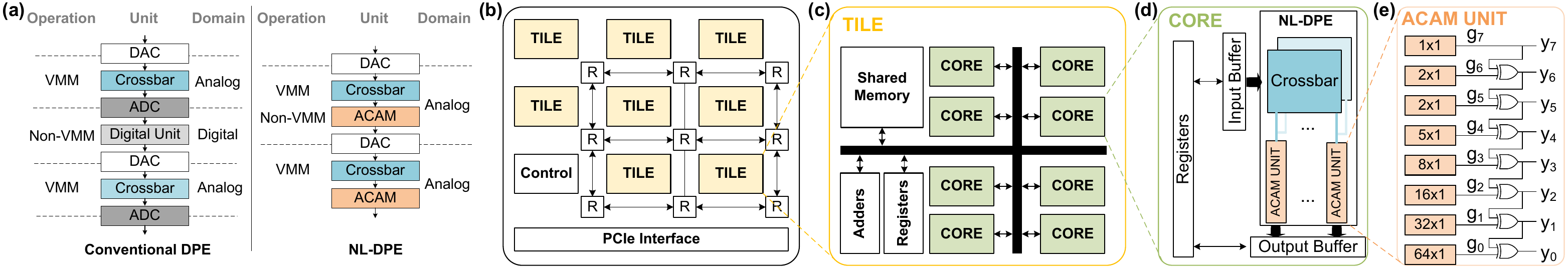}
\caption{\textbf{(a)} Conceptual representation of conventional DPE (left) and NL-DPE (right), which replaces ADC and digital computing logic. \textbf{(b)} Schematic of the accelerator tiled architecture. \textbf{(c)} Tile schematic. \textbf{(d)} Core schematic. \textbf{(e)} ACAM unit schematic.}
\vspace{-0.2in}
\label{fig:overview}
\end{center}
\end{figure*}

While the Linear layers can be efficiently implemented on RRAM crossbars, there are two design challenges:
\begin{itemize}[noitemsep,nolistsep,leftmargin=*]
\item DMMul computation. Given $\mathbf{Q}$, $\mathbf{K}$, and $\mathbf{V}$ are dynamically generated for each new input, accelerating DMMul with conventional DPE~\cite{yang2020retransformer,yazdanbakhsh2022sparse} demands programming RRAM crossbars frequently, as shown in Fig.~\ref{fig:attention}(b), which may significantly degrade accelerator's energy efficiency and RRAM cells' endurance.
An alternative is to compute DMMuls in digital circuits using VFUs~\cite{ankit2019puma}, as shown in Fig.~\ref{fig:attention}(c), which becomes the dominant energy consumer as reported in Fig.~\ref{fig:breakdown}.

\item Softmax computation. As shown in Equation~\ref{eq:softmax}, Softmax involves exponentiation and division, both of which are difficult to implement using RRAM crossbars. While Softmax computation is typically performed using digital units, it leads to degraded energy efficiency as reported in Fig.~\ref{fig:breakdown}.
\end{itemize}

\noindent
Notably, since both DMMul and Linear layers only contain linear computations, the scaling (i.e., division by $\sqrt{d_k}$) can be fused into the weights (i.e., $\mathbf{W_Q}$, and $\mathbf{W_K}$) during inference.

\section{Non-Linear Dot-Product Engine}

\subsection{Overview}

Fig.~\ref{fig:overview}(a) illustrates our key innovation by comparing the implementations of a snippet of a DNN using a conventional DPE architecture~\cite{shafiee2016isaac} and the proposed NL-DPE, respectively.
The snippet consists of three layers: two VMM-based layers (e.g., Convolutional layer) and one non-VMM layer (e.g., activation layer).
The conventional DPE performs non-VMM operations digitally, and thus requires ADCs and DACs to convert analog output from the crossbar before non-VMM operations and convert non-VMM output to analog signals before the crossbar of the next layer, respectively.
By contrast, the proposed NL-DPE performs non-VMM operations directly in the analog domain using ACAMs, eliminating the need for ADCs after the crossbar, as the ACAMs can process analog inputs directly and output a digital value.
While DACs are still necessary before subsequent VMM layers, they consume significantly less energy and area than ADCs~\cite{shafiee2016isaac}.
For VMM layers not followed by non-VMM operations, the ACAMs compute an identity function, effectively acting as ADC substitutes and enabling the entire system to operate without ADCs.

We next present an overview of the NL-DPE architecture in Fig.~\ref{fig:overview}(b)-(e).
An NL-DPE accelerator chip consists of multiple tiles organized as a 2D mesh with tiles connected using on-chip routers, orchestrated by a controller. This is similar to the popular tile-core accelerator hierarchy~\cite{ankit2019puma}, which communicates with the host system via a PCIe interface.

Each tile includes a shared memory for storing the input and output of the computations assigned to it, along with a set of adders and associated registers to accumulate partial results from the cores shown in Fig.~\ref{fig:overview}(d).
Each core contains an NL-DPE composed of multiple crossbars and ACAM units.
One ACAM unit computes the output from the same column of all the crossbars in the same core, while one ACAM unit contains multiple ACAM arrays, with each array computing one bit of the output, as shown in Fig.~\ref{fig:overview}(e).
XOR gates are used in the ACAM unit to decode its output for subsequent computations.
\subsection{Computation Modes}

Given that NL-DPE computes VMM operations in crossbars and non-linear functions in ACAMs, it can be configured into one of the following three modes: 

\textbf{(1) Dual-compute mode}.
This is the typical configuration as a DNN accelerator. i.e., the crossbars are configured to compute VMMs while the ACAMs are configured to compute non-linear functions. 
The crossbars store the weights of the Convolutional or Linear layers and compute the dot-product between the weights and the input, while the ACAMs store the thresholds of the DTs, which are trained to approximate the target non-linear function (Section~\ref{sec:act}). 
A similar mechanism can be used to compute the attention mechanism by converting the Linear layer outputs in the logarithmic domain (Section~\ref{sec:complicate_functions}).

\textbf{(2) Crossbar-only mode}.
In this configuration, NL-DPE only needs to perform a VMM within the crossbars, e.g., in the case when a Convolutional layer is followed by another Convolutional layer rather than a non-linear activation layer. We then configure ACAMs to compute the identity function, which effectively acts as conventional ADCs, converting the analog output from the crossbars into digital signals.

\textbf{(3) ACAM-only mode}.
In this configuration, NL-DPE only exploits the ACAMs to compute non-linear operations. Configuring a subset of NL-DPE cores in this mode enables the computation of more complicated non-linear functions. In this configuration, the crossbars store identity matrices such that their inputs are directly passed to the ACAMs for non-linear operations. That is, NL-DPE functions as a vector ALU, which aims to replace VFUs.

By altering the contents stored in the RRAM cells of crossbars and ACAMs, NL-DPE can flexibly switch all or a subset of cores to different modes. No hardware modification is needed after deployment.  Thus, the NL-DPE itself is agnostic to the selected mode.

\begin{figure}[ht]
\begin{center}
\includegraphics[width=0.8\linewidth]{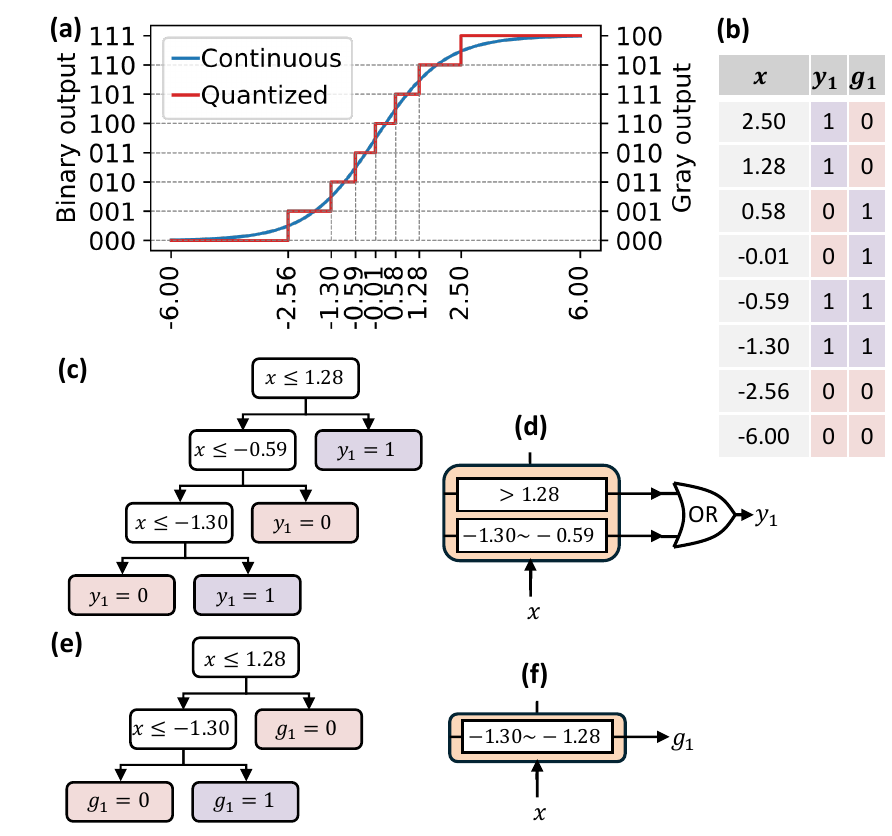}
\caption{\textbf{(a)} The Sigmoid function with output being quantized to 3 bits in unsigned binary ($y$ on the left Y axis) and in Gray code format ($g$ on the right Y axis). \textbf{(b)} Training dataset for predicting $y_1$ and $g_1$. \textbf{(c)} Trained DT to predict $y_1$. \textbf{(d)} Mapping (c) into an ACAM. \textbf{(e)} Trained DT to predict $g_1$. \textbf{(f)} Mapping of (e) into an ACAM.}
\vspace{-0.1in}
\label{fig:activation}
\end{center}
\end{figure}

\subsection{Computing simple non-linear functions with ACAMs}\label{sec:act}

We next use Sigmoid as an example to demonstrate how to configure ACAMs to compute a non-linear single-variable function.
Fig.~\ref{fig:activation}(a) shows the output of the Sigmoid function when it is quantized to three bits.
The blue and red lines represent the Sigmoid function in continuous and quantized versions, respectively.
In this example, we assume that 0 and 1 are quantized as $000_b$ and $111_b$, respectively, with the remaining 6 values evenly distributed between 0 and 1, as shown by the left Y-axis in the figure.
Note that this simple quantization scheme is only for illustration purposes, an arbitrary quantization scheme can be used.

We convert the computation of each output bit $y_i$ (0$\le$i$\le$2) to a binary classification problem and predict its value using one DT. 
For the example, the $y_1$ column in the table in Fig.~\ref{fig:activation}(b) tracks its value change at different thresholds, e.g., 
(x=1.28, y$_1$=1) and (x=0.58, y$_1$=0) indicate y$_1$ being 1 if x$>$1.28; and 0 if x$\le$1.28.
The complete DT for $y_1$ is shown in Fig.~\ref{fig:activation}(c). Unlike traditional DTs that predict unseen data, our binary-bit DTs are specifically designed to memorize the exact thresholds in the training dataset, intentionally overfitting the data for precise computation.

Fig.~\ref{fig:activation}(d) shows how the resulting DT is mapped onto an ACAM array.
Since the tree has only one feature, the ACAM array logically contains a single column, with the number of rows corresponding to the number of leaf nodes where $y_1$=1 (two rows in this example).
All match lines in the ACAM array are connected to an OR gate, which outputs the computed result for $y_1$.
In practice, multiple ACAM cells can be arranged in a row to perform the OR operation through a pass transistor logic, significantly reducing the logic gate overhead.
If the input falls within any of the ranges stored in the two ACAM cells, $y_1$ is set to 1.
Unlike the Look-Up Table (LUT) approach, where an attached memory is needed to retrieve the output (see the RAM after the ACAM array in Fig.~\ref{fig:dpe_acam}(d))~\cite{zhu2022fuse}, the ACAM's output directly represents the bit value, eliminating the need for additional memory access.
This method can be applied to compute the other bits ($y_0$ and $y_2$), requiring a total of three ACAM arrays to calculate the 3-bit Sigmoid function in this example.

\noindent
{\underline{\bf Encoding Optimization.}}\label{sec:encoding}
From Fig.~\ref{fig:activation}(b) and (d), we can see that the number of rows in each ACAM array for computing a single output bit is determined by the number of consecutive 1s in the training dataset, i.e., how frequently the output bit toggles between 0 and 1 as the input value increases.
Since activation functions are typically monotonic, the less significant output bits tend to toggle more frequently between 0 and 1.
For an $n$-bit activation function, there are $2^n$ distinct output levels, meaning the worst-case scenario for the number of ACAM rows is $2^{n-1}$.

To reduce ACAM array size, we propose an encoding scheme based on Gray code to minimize the bit-toggling rate.
Originally designed to minimize error probabilities in analog-to-digital signal conversion, Gray code ensures that only one-bit changes occur at a time between consecutive numbers.
For example, considering 3 bits, when the decimal value increases from 3 to 4, there are three-bit flips for binary format ($011_b \rightarrow 100_b$), while the Gray code has only one-bit flip ($010_g \rightarrow 110_g$).

In Fig.~\ref{fig:activation}(a) (right Y-axis), we show the output of the example in Gray code.
Instead of computing $y_1$, we compute $g_1$ instead.
Fig.~\ref{fig:activation}(e) and (f) show the trained DT and its corresponding ACAM mapping.
The number of ACAM rows is reduced from 2 to 1.

Table~\ref{tbl:activations} compares the required ACAM sizes for various 8-bit activation functions predicted with DTs using unsigned binary (B) or Gray coding (G), demonstrating that the latter reduces the ACAM size for all output bits (except the MSB) by approximately half.
{\ccblue
The last row in the table shows a negligible Mean Squared Error (MSE) compared with digital computation, indicating that the basic function computations incur negligible accuracy loss.
}

\begin{table}[ht]
\centering
\caption{Number of ACAM rows for various 8-bit functions.}
\scriptsize 
\label{tbl:activations}
\resizebox{0.45\textwidth}{!}{%
\begin{tabular}
{|c|c|c|c|c|c|c|c|c|}
\hline
& \textbf{Sigmoid} & \textbf{Tanh} & \textbf{SiLU} & \textbf{GELU} & \textbf{ReLU} & \textbf{Identity} & \textbf{log} & \textbf{exp}\\
\hline
\textbf{Enc} & B/G & B/G & B/G & B/G & B/G & B/G & B/G & B/G \\
\hline
\textbf{bit$_7$}&1/1   &1/1   &1/1   &1/1   &1/1   &1/1  &1/1  &1/1\\
\textbf{bit$_6$}&2/1   &2/1   &2/1   &2/1   &2/1   &2/1  &3/2  &2/1\\
\textbf{bit$_5$}&4/2   &4/2   &4/2   &4/2   &4/2   &4/2  &5/2  &4/2\\
\textbf{bit$_4$}&8/4   &8/4   &8/4   &8/4   &8/4   &8/4  &8/5  &8/4\\
\textbf{bit$_3$}&16/8  &16/8  &16/8  &16/8  &16/8  &16/8 &16/8 &16/8\\
\textbf{bit$_2$}&32/16 &32/16 &32/16 &32/16 &32/16 &32/16&32/16&32/16\\
\textbf{bit$_1$}&64/32 &64/32 &64/32 &64/32 &64/32 &64/32&64/32&64/32\\
\textbf{bit$_0$}&121/64&113/64&101/64&112/64&121/64&1/64 &97/64&108/64\\
\hline
\textbf{Total}&248/128&240/128&228/128&239/128&248/128&128/128&226/130&235/128\\
\hline
{\ccblue \textbf{MSE}}&{\ccblue 6.15e-08}&{\ccblue 6.15e-07}&{\ccblue 7.01e-23}&{\ccblue 2.46e-06}&{\ccblue 2.05e-06}&{\ccblue 1.29e-05}&{\ccblue 3.03e-06}&{\ccblue 3.72e-06}\\
\hline
\end{tabular}
}
\end{table}

Gray code, while reducing the number of bit flips, cannot be directly involved in binary computation. 
To enable subsequent computations, we convert the Gray code outputs from the ACAMs back to their binary form.
This is achieved as follows.
\begin{equation*}
y_i = \begin{cases}
g_i &i=n-1\\
XOR(g_{n-1}, g_{n-2}, ..., g_i) &i<n-1
\end{cases}
\end{equation*}
where $y_i$ and $g_i$ are the $i$th bit in the binary and Gray code format.
$n$ is the number of output bits.

\noindent
{\underline{\bf ACAM unit implementation.}}
The detailed structure of an ACAM unit is illustrated in Fig.~\ref{fig:overview}(e). 
We selected a bit-width of 8 after careful design space exploration, as shown in Section \ref{sec:tradeoff-bitwidth}, thus each ACAM unit consists of 8 ACAM arrays, with each array responsible for computing one output bit.
The sizes of these 8 ACAM arrays are determined based on the largest requirement across all functions profiled in our tested AI models. (Table~\ref{tbl:activations}).
Specifically, starting from the most significant output bit, the sizes of the 8 ACAM arrays are 1, 2, 2, 5, 8, 16, 32, and 64.
After the ACAM arrays, we use 7 XOR gates for decoding Gray code to binary form.

\subsection{Computing more complicated functions with ACAMs} \label{sec:complicate_functions}
For more complicated non-linear functions, such as Softmax and DMMul,
we observe that deploying one NL-DPE cannot compute their output bits efficiently.
We next propose methods to configure NL-DPE adaptively to compute these functions. 

\noindent
{\underline{\bf Data-Dependent Matrix Multiplication (DMMul).}} 
A major difference between Sigmoid and DMMul is that DMMul depends on two inputs, both of which change dynamically. A naive implementation of DMMul in NL-DPE demands writing one input into RRAM cells, which not only introduces prohibitive performance and energy overheads but also degrades RRAM lifetime dramatically. 

\begin{figure*}[tbp]
\centerline{\includegraphics[width=\linewidth]{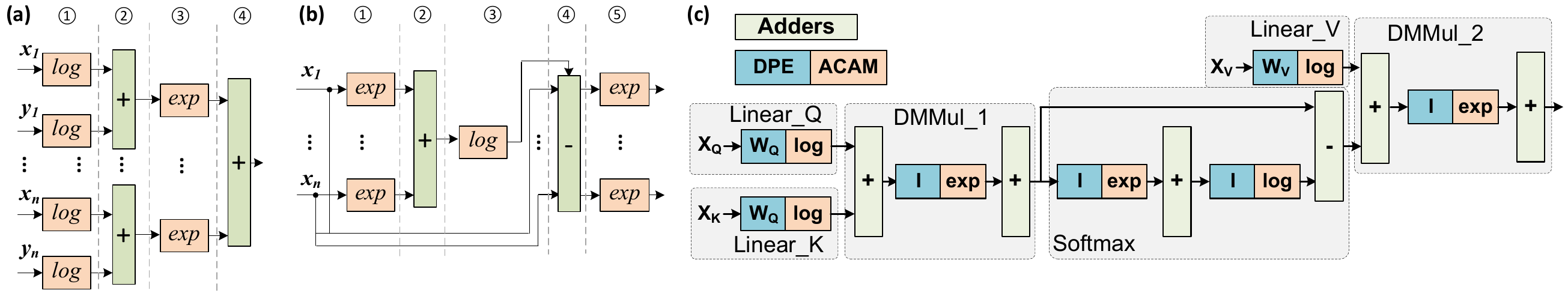}}
\caption{\textbf{(a) }Steps of computing the dot-product between two vectors after replacing multiplication with exponentiation, logarithm, and addition. \textbf{(b)} Steps of computing Softmax with exponentiation, logarithm, and addition. \textbf{(c) }Mapping the attention mechanism onto NL-DPEs.}
\vspace{-0.2in}
\label{fig:attention_mapping}
\end{figure*}

To address this, we propose to convert each multiplication into a sequence of logarithmic, exponential, and addition operations, i.e.:
\begin{equation}
a \times b = e^{\left(\log(a)+\log(b)\right)}
\end{equation}
Both exponentiation and logarithm functions are single-variable operations, which are suitable for computation using ACAMs in the NL-DPE.
Fig.~\ref{fig:attention_mapping}(a) illustrates the steps of computing the dot-product between two vectors, a fundamental component of DMMul.
In Step 1, an element-wise logarithm is computed for both input vectors.
With all elements now on the log scale, Step 2 replaces the multiplications with additions.
In Step 3, the result is converted back to the linear scale through element-wise exponentiation.
Finally, the linear scale values are summed to produce the final output.

In the figure, the logarithm and exponential operations are implemented using the ACAMs, while the add operations use the on-chip adders.

\noindent
{\underline{\bf Softmax.}}
Softmax entails exponentiation and division operations, both of which incur high digital hardware costs.
By studying the Softmax function in Equation~\ref{eq:softmax}, we identify that its expensive exponentiation and division operations can be easily mitigated with ACAM-based computing. In particular, we exploit logarithm operations to convert a division operation as follows~\cite{yang2020retransformer}:
\begin{equation}
\frac{a}{b} = e^{\left(\log(a)-\log(b)\right)}
\end{equation}

Fig.~\ref{fig:attention_mapping}(b) illustrates the steps of computing the Softmax function after the conversion.
In Step 1, the exponentiation of each element in Softmax’s input vector is computed.
Next, these exponentiated values are summed to obtain the denominator in Equation~\ref{eq:softmax} (Step 2).
In Step 3, this sum is converted to the log scale to use the subtraction in Step 4 and substitute the original division.
Finally, Step 5 converts the result back to the linear scale with element-wise exponentiation.

\noindent
{\underline{\bf Attention mechanism and its NL-DPE implementation.}}
We next use Fig.~\ref{fig:attention_mapping}(c) to illustrate how to configure multiple NL-DPEs to implement complicated non-linear functions. We use the attention mechanism in Fig.~\ref{fig:attention} as an example.

DMMul and Softmax extensively rely on exponentiation and logarithm operations, both of which can be implemented using the NL-DPE in ACAM-only mode.
By examining the entire attention computation flow in Fig.~\ref{fig:attention}, we can combine certain computations to optimize efficiency.
Specifically, the inputs to the first DMMul are generated by Linear layers, and the output of Softmax feeds into the second DMMul.
Fig.~\ref{fig:attention_mapping}(c) illustrates an optimized mapping of the attention mechanism that minimizes redundant computations.
Since the two inputs to DMMul\_1 and the second input to DMMul\_2 are produced by Linear layers, we can simplify by using a single NL-DPE to handle both the Linear layers and the logarithm needed for DMMuls, treating the logarithm as an activation following the Linear layers.
Additionally, because the final step in Softmax involves an element-wise exponentiation and the first step in DMMul\_2 involves a logarithm—operations that are inverses of each other, these steps can be bypassed.
As shown in Fig.~\ref{fig:attention_mapping}(c), the log-scale output of Softmax is fed directly into DMMul\_2, which no longer requires logarithmic processing for its second input.
For all other computations, we still use NL-DPEs in the ACAM-only mode by storing an identity matrix in its crossbars.

\section{Addressing Non-Idealities in Analog IMC}

In this section, we present our noise-aware fine-tuning (NAF) approach to enhance the noise resilience of NL-DPE.
We first present our noise model and then elaborate on the NAF details.

\subsection{Noise Models}\label{sec:noise}
It is well-known that RRAM-based IMC accelerators suffer from conductance noises, i.e., 
there is a non-negligible difference between the desired cell conductance and the cell conductance that was actually programmed or can be read \cite{mao2022experimentally} during computing. The noises come from two main sources.
\begin{itemize}[noitemsep,nolistsep,leftmargin=*]
    \item A RRAM cell may not be precisely programmed to the desired conductance levels. This is due to the stochastic nature of the ionic migration during the programming operation.
    \item Each read operation from a programmed memory cell may yield slightly varying conductance values. This fluctuation is due to various physical phenomena, such as thermal and random telegraph noise.
\end{itemize}
RRAM demonstrates minimal conductance drift after programming, which is neglected in this paper.

\begin{figure}[ht]
\begin{center}
\includegraphics[width=\linewidth]{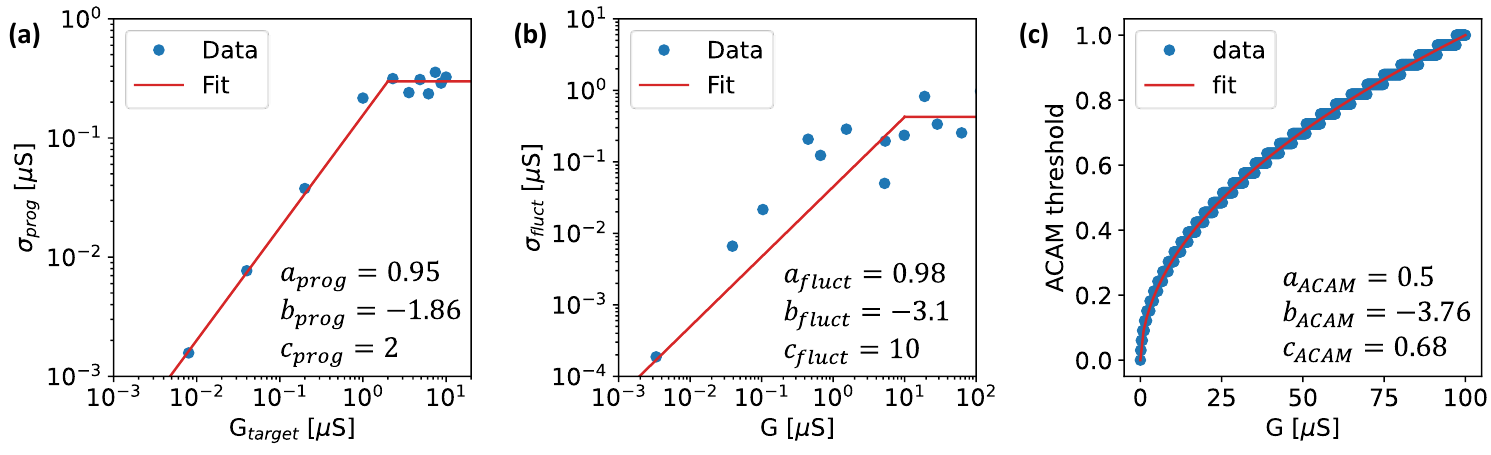}
\caption{Experimental data and model fitting of \textbf{(a)} standard deviation of programmed conductance $\sigma_{prog}$ as a function of the target conductance $G_{target}$, \textbf{(b)} standard deviation of fluctuated conductance $\sigma_{fluct}$ as a function of the mean conductance $G$, and \textbf{(c)} transfer function from the programmed conductance to ACAM threshold.}
\vspace{-0.2in}
\label{fig:noise}
\end{center}
\end{figure}

In this paper, we develop a noise model by fitting it to noise data collected from a fabricated RRAM test chip~\cite{sheng2019low}. 
We run a program-and-verify algorithm on Ta-Ox RRAM devices using a tolerance of $\pm0.55\mu$S for $G_{target}>1\mu$S and tolerance proportional to the conductance levels for $G_{target}\leq1\mu$S. We then read the conductance values 1000 times to assess the fluctuation noise.

Fig.~\ref{fig:noise}(a) indicates that the maximum standard deviation of programming noise $\sigma_{prog}$ is approximately 0.4$\mu$S and can be modeled as a function of the target conductance $G_{target}$. 
Fig.~\ref{fig:noise}(b) indicates that the standard deviation of read fluctuation $\sigma_{fluct}$ can be modeled as a function of the mean conductance $G$.
These experimental results also match those reported in \cite{mao2022experimentally}, where the log-scale standard deviation increases approximately linearly with the mean until it saturates at a certain point, which may be related to the threshold between low and high conductance states.

We model the programming and read noises using a normal distribution and compute their standard deviation as follows.
\begin{equation}
\sigma_x = \text{exp}(a_x\text{log}(G.\text{clip}(0,c_x)+b_x)
\end{equation}
We fit the parameters $a_x$, $b_x$, and $c_x$ to the experimental data and report them in Fig.~\ref{fig:noise}(a,b).
The programming and read fluctuation errors are then computed as $G_{write}=\sigma_{prog}\cdot \mathcal{N}(0,1)$ and $G_{read}=\sigma_{fluct}\cdot \mathcal{N}(0,1)$, respectively, where $\mathcal{N}(0,1)$ is a normal distribution with mean 0 and standard deviation of 1.

Given a RRAM cell that was programmed to its target conductance $\mathbf{G}_{target}$, we compute the readout conductance $G$ as follows.
\begin{equation}\label{eq:g_model}
\mathbf{G} = \mathbf{G}_{target} + \mathbf{G}_{write} + \mathbf{G}_{read}.
\end{equation}
The conductance model in Eq.~\ref{eq:g_model} applies to RRAM cells in both crossbars and ACAM units. 
However, ACAM involves a non-linear operation, so the relation between conductance and the ACAM threshold needs to be modeled as well.
Fig.~\ref{fig:noise}(c) reports the experimental data on the ACAM threshold values as a function of the programmed conductance~\cite{pedretti2021tree}.
We then model the transfer function as 
\begin{equation}
    \text{TH} = \text{exp}(a_{ACAM}*\text{log}(G)+b_{ACAM})+c_{ACAM}
\end{equation}
with the fitting parameters reported in Fig.~\ref{fig:noise}(c).

Note, we did not consider thermal and PVT variations in the paper. Their impacts on ACAM stability have been studied in prior works~\cite{pedretti2022differentiable}. The corresponding solutions, e.g., opportunely programming the RRAM devices to address the process variations on threshold voltage in the input transistors, are applicable because NL-DPE programs each ACAM individually, without the integration in the full model.

\subsection{Noise-Aware Fine-tuning (NAF)} \label{sec:dpe_noise}

\begin{figure}[ht]
\begin{center}
\includegraphics[width=3.3in]{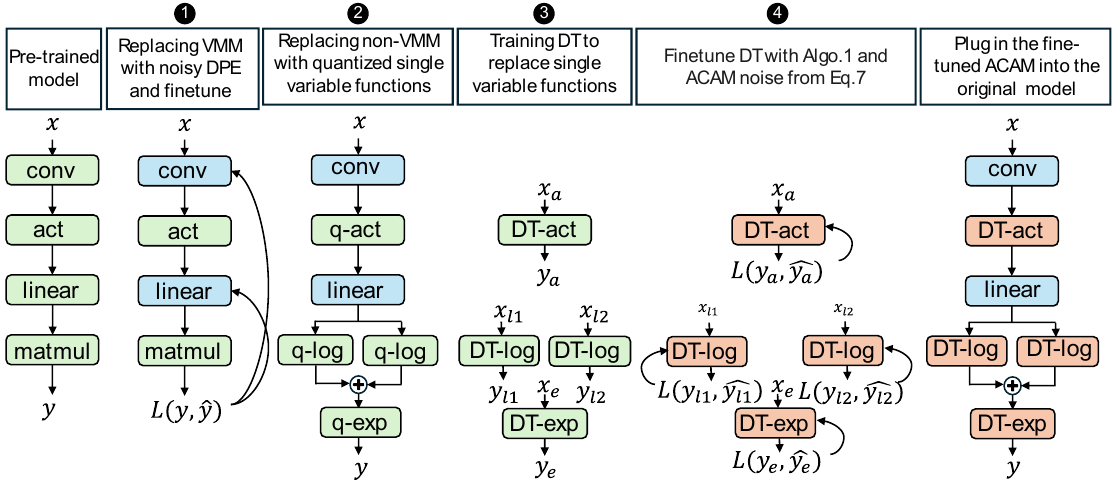}
\caption{Steps of mitigating noises in crossbars and ACAMs.}
\vspace{-0.1in}
\label{fig:finetune}
\end{center}
\end{figure}

NL-DPE adopts Noise-Aware Fine-tuning (NAF) to achieve noise-resilience. As shown in Fig. \ref{fig:finetune}, NAF is a software-based per-DT approach such that, given a pre-trained AI model, NAF extracts weight parameters and non-VMM operations, which shall be programmed to crossbars and ACAMs, respectively. We convert non-VMM operations to their corresponding DTs. NAF fine-tunes the model in software before deployment by integrating the noise model in both crossbars and ACAMs. While a few end-to-end model fine-tuning rounds are needed for crossbars, each DT is fine-tuned independently, without being integrated in end-to-end model fine-tuning. The fine-tuned model is then programmed to RRAM cell arrays, requiring no post-deployment in-device fine-tuning.

Fig.~\ref{fig:finetune} illustrates the overall process. 
Given a pre-trained AI model, Step \ballnumber{1} addresses noises in the crossbars, which is accomplished by a small number of NAF iterations (typically fewer than 10). 
During each iteration, NAF injects random noise based on our noise model (Eq.~\ref{eq:g_model}) into the weights of Convolutional and Linear layers.

Step \ballnumber{2} converts all non-VMM operations (such as DMMul and Softmax) into single-variable functions (e.g., log, exp, etc.).
The output of each single-variable function is quantized so that it is ready for training DTs in Step \ballnumber{3}.
Step \ballnumber{3} trains the DTs for each non-VMM operation that will be eventually deployed on ACAMs.

Step \ballnumber{4} addresses ACAM noises that may impact DT's accuracy by performing NAF iterations for each DT \textit{independently}.
This is different from Step~\ballnumber{1} such that it does not require end-to-end model fine-tuning. After all these steps, the fine-tuned DTs replace the non-VMM operations in the original AI, which is then deployed to device by programming RRAM cells and is ready to use. 
We next discuss Steps \ballnumber{1} and \ballnumber{4} in more details.

{\underline{\bf Mitigating crossbar noise.}}
Most prior works adopt digital slicing (D-SL) \cite{shafiee2016isaac} to map model weights onto RRAM cells in crossbars.
However, D-SL is suboptimal for mitigating RRAM noise, as it stores \textit{discrete} programmed values in each cell.
In contrast, we employ analog slicing (A-SL) \cite{pedretti2021redundancy}, which maps model weights to RRAM cells using \textit{continuous} analog values.
This enables us to propose a novel A-SL-aware loss function for NAF to effectively reduce the impact of RRAM noise in crossbars.

\begin{figure}[ht]
\begin{center}
\includegraphics[width=0.8\linewidth]{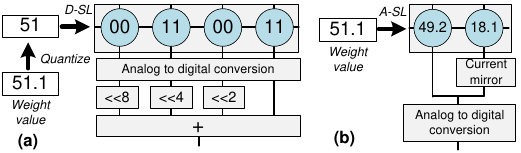}
\caption{Mapping a weight value onto RRAM cells in a crossbar with \textbf{(a)} digital slicing and \textbf{(b)} analog slicing.}
\vspace{-0.1in}
\label{fig:slicing}
\end{center}
\end{figure}

Figure~\ref{fig:slicing} illustrates the differences between D-SL and A-SL.
In D-SL, a full-precision weight is first quantized to a fixed-point representation (e.g., 8 bits), with each bit segment stored in a separate RRAM cell.
During computation, a shift-and-add circuit is used to combine the outputs from individual cells.
In contrast, A-SL directly programs each RRAM cell’s conductance to reflect a continuous analog value corresponding to the target weight.
However, due to limited programming precision, there exists a residual error (denoted as $\epsilon$) between the programmed conductance and the desired value.
To compensate for this error, a second RRAM cell is used to store a scaled version of the error (e.g., $10\times\epsilon$).
During inference, an analog current mirror scales down the second cell’s output before combining it with the output of the first cell.

Because A-SL stores continuous values in each RRAM cell, we can incorporate the inter-cell error ($\epsilon$) as a regularization term in the NAF loss function:
\begin{equation}
Loss = MSE(y, \hat{y}) + \lambda_1 \|W\|_\infty + \lambda_2 \|\epsilon\|_\infty
\end{equation}
Here, we use Mean Squared Error (MSE) between the model output and ground truth as the primary loss function.
We also add an $L_\infty$ regularization on the weights $W$ to encourage smaller target conductance values, which are known to be less susceptible to noise according to Fig.~\ref{fig:noise}(a) and (b).
The coefficients $\lambda_1$ and $\lambda_2$ control the strength of these regularizations.

\begin{algorithm}[tbp]
  \footnotesize
  \caption{Differentiable Approximation of ACAM}
  \label{algo:naf}
  \SetAlgoLined
  \KwIn{$x$: one input value for fine-tuning;
        $\mathbf{w}_{i}^{L}$: lower thresholds of the DT for $i$th output bit; 
        $\mathbf{w}_{i}^{H}$: higher thresholds of the DT for $i$th output bit;}
  \KwOut{$y$: an 8-bit output of the ACAM} 
  \KwData{$g_{ratio}$: weight-to-conductance ratio; 
          $g_{min}$: minimum conductance of RRAM; 
          $g_{max}$: maximum conductance of RRAM; 
          $\epsilon$: a very small positive constant to avoid division-by-zero.}

  \For{$i = 0$ \KwTo $7$}{
      $\mathbf{g}_{i}^{L} = \mathrm{Clip}(\mathbf{w}_{i}^{L} \cdot g_{ratio} + g_{min}, [g_{min}, g_{max}])$\;
      $\mathbf{g}_{i}^{H} = \mathrm{Clip}(\mathbf{w}_{i}^{H} \cdot g_{ratio} + g_{min}, [g_{min}, g_{max}])$\;
      $\mathbf{\tilde{g}}_{i}^{L} = \mathrm{Noise}(\mathbf{g}_{i}^{L})$;
      $\mathbf{\tilde{g}}_{i}^{H} = \mathrm{Noise}(\mathbf{g}_{i}^{H})$\;
      $\mathbf{\tilde{w}}_{i}^{L} = (\mathbf{\tilde{g}}_{i}^{L} - g_{min}) / g_{ratio}$\;
      $\mathbf{\tilde{w}}_{i}^{H} = (\mathbf{\tilde{g}}_{i}^{H} - g_{min}) / g_{ratio}$\;
      $\mathbf{m}_{i} = \mathrm{ReLU}(x - \mathbf{\tilde{w}}_{i}^{L}) \cdot \mathrm{ReLU}(\mathbf{\tilde{w}}_{i}^{H} - x)$\;
      $m_{i} = \mathrm{Sum}(\mathbf{m}_{i})$\;
      $m_{i} = m_{i} / (m_{i} + \epsilon)$\;
  }

  $y = 0$\;
  \For{$i = 7$ \KwTo $0$}{
      \uIf{$i == 7$}{
          $b_{i} = m_{i}$\;
      }
      \Else{
          $b_{i} = (m_{i} - b_{i+1})^2$\;
      }
      $y = y + b_{i} \cdot 2^i$\;
  }
\end{algorithm}

{\underline{\bf Mitigating ACAM noise.}}
To fine-tune the DT thresholds for tolerating analog noises, we incorporate the DT into a training process that includes a forward pass and a backward pass, similar to a neural network.
This process requires computing the gradients of the DT thresholds for updates.
However, the computations in DTs, such as the comparisons between inputs and thresholds, are inherently non-differentiable, making gradient computation infeasible.
To address this challenge, we propose a differentiable approximation of the DT computation based on its implementation in ACAM~\cite{pedretti2022differentiable, zhao2024noise, zhao2023race}.

Algorithm~\ref{algo:naf} demonstrates the differentiable method for computing an 8-bit output using 8 DTs mapped to ACAMs.
After extracting noisy DT thresholds from the conductance models (Line 2-6), in line 7, we replace the comparison operation between inputs and thresholds with a ReLU function.
This ensures that $\mathbf{m}_{i}$ is positive if the input $x$ lies within the range defined by the lower and upper thresholds.
Line 8 uses a sum operation as a differentiable replacement for the OR gate in ACAM.
Line 9 employs a division to quantize ${m}_{i}$ to 0 or 1, with a small $\epsilon$ added to prevent division by zero.
After the first loop, each ${m}_{i}$ is a floating-point number very close to either 0 or 1, representing the Gray code output bit of an ACAM array.
The second loop, in lines 12–19, implements a differentiable version of the XOR-based decode logic, replacing it with a combination of subtraction and squaring to maintain differentiability.

As a result, all computations in Algorithm~\ref{algo:naf} are fully differentiable, allowing gradients of the threshold tensors to be computed during backpropagation.

\section{Methodology} \label{sec:methodology}
Key parameters, area, and power of NL-DPE components are summarized in Table~\ref{tab:architecture}.
Most parameters are adopted from ISAAC except crossbars and ACAMs, which are modeled via 16nm SPICE simulations~\cite{li2020analog}.
For an 86$\times$12 ACAM array, energy is measured by integrating voltage and current with all MLs discharging in the worst case.
Each ACAM cell occupies 0.72$\mu m^2$, consumes $\sim$0.44$fJ$ per search, and completes a search in $\sim$300$ps$.
Each tile has 8 cores, each with 4 crossbars (two for positive and two for negative weights, size $256 \times 256$), with each column connected to an ACAM unit containing 8 ACAM arrays.
A total of 130 ACAM cells per unit supports all activation functions (Table~\ref{tbl:activations}).
The RRAM conductance range was set to 6.7$k\Omega$/100$M\Omega$ (i.e., 0.01$\mu$S - 150$\mu$S) for optimal energy-accuracy trade-off (Section~\ref{sec:tradeoff-conductance}).
Noise models (Eq.~\ref{eq:g_model}) are applied at different NAF stages (Fig.~\ref{fig:finetune}), validated with data from a fabricated chip~\cite{sheng2019low}.

We compare NL-DPE with three baselines: (1) NVIDIA H100 GPU; (2) CMOS systolic array accelerator Eyeriss~\cite{chen2016eyeriss}; and (3) RRAM-based IMC accelerators ISAAC~\cite{shafiee2016isaac}, RAELLA~\cite{andrulis2023raella}, and BRAHMS~\cite{song2021brahms}.
ISAAC uses ADCs for all VMM outputs and VFUs for non-VMM operations.
RAELLA reduces ADC overhead via adaptive bit slicing but still relies on VFUs, while BRAHMS integrates ACAM for ReLU and hard-Tanh activations, requiring larger ACAM arrays and separate ADC/activation units.
To support arbitrary activations, we augment all non-GPU baselines with a piecewise-linear Flex-SFU~\cite{reggiani2023flex}, which is only invoked for activations not originally supported (e.g., GeLU in BERT, SiLU in EfficientNet).
Flex-SFU incurs higher energy than optimized digital logic.
All accelerators are assumed to have enough tiles for 100M-parameter models, except in the NL-DPE multi-chip scalability analysis (Section~\ref{sec:scalability}), where the die area is matched to an H100 GPU.

Our GPU baseline is the NVIDIA H100 (16896 CUDA cores, 528 Tensor Cores, 80GB HBM3).
All models are quantized to INT8 to match NL-DPE and exported in ONNX format.
Quantization and inference use TensorRT, with 200 warm-up iterations followed by 1000 runs for measurement.
Latency is reported by TensorRT, average power via NVIDIA-SMI, and accuracy is evaluated using floating-point GPU outputs as a high-quality baseline.

NL-DPE and all baselines are simulated in 32nm using CiMLoop~\cite{andrulis2024cimloop}.
For ISAAC and RAELLA, we use CiMLoop’s implementations~\cite{shafiee2016isaac,andrulis2023raella}, which originally support only VMMs.
We extend it to non-VMMs by mapping the first operand as input and the second as weight, adding a Flex-SFU~\cite{reggiani2023flex} and specifying operand mappings for scheduling.
Single-operand operations use a one-element weight.
DMMul is modeled as a grouped convolutional layer, and element-wise operations as a special convolutional layer, whose input is of size $N\times 1 \times 1 \times 1$ (i.e., $N$ images, each image has only 1 pixel), where $N$ is the element-wise operation size.
Memory access and on-chip network overhead are included.
For BRAHMS and NL-DPE, ADC and Flex-SFU are replaced with ACAM, implemented as custom CiMLoop plugins.

\begin{table}[ht!]
\centering
\caption{NL-DPE parameters.}
 \scriptsize
\label{tab:architecture}
\resizebox{1.0\linewidth}{!}{%
\begin{tabular}{|c|c|c|c|c|}
\hline
\multicolumn{5}{|c|}{\textbf{1GHz on 32nm Technology node}} \\
\hline
\textbf{Component} & \textbf{Spec} & \textbf{Params} & \textbf{Power} ($mW$) & \textbf{Area} ($mm^2$)\\
\hline
\multicolumn{5}{|c|}{Core (8 cores per tile)} \\
\hline
DPE & \begin{tabular}{c} number \\ size \end{tabular} & \begin{tabular}{c} 4 \\ $256 \times 256$ \end{tabular} & 1.31 & 0.011534 \\
\hline
ACAM & \begin{tabular}{c} number \\ size \end{tabular} & \begin{tabular}{c} 256 \\ $130 \times 1$ \end{tabular} & 43.52 & 0.041431 \\
\hline
Input buffer & size & 256B & 0.23 & 0.00077 \\
\hline
Output buffer & size & 256B & 0.23 & 0.00077 \\
\hline
Register & size & 128B & 0.12 & 0.000385 \\
\hline
DAC & number & $4 \times 256$ & 4 & 0.00017 \\
\hline
XOR & number & 7$\times$256 & 0.385 & 0.000215 \\
\hline
Core Total & - & - & 49.795 & 0.055275 \\
\hline
\hline
\multicolumn{5}{|c|}{Tile} \\
\hline
Register & size & 768B & 0.69 & 0.00231 \\
\hline
Adders & number & 256 & 12.8 & 0.0154 \\
\hline
Shared memory & data & 64KB & 20.7 & 0.083 \\
\hline
Core & number & 8 & 398.36 & 0.4422 \\
\hline
Tile Total & - & - & 432.55 & 0.54291 \\
\hline
\end{tabular}
}
\end{table}

\begin{table*}[ht]
\caption{Accuracy of various stages in the proposed NAF.}
\label{tbl:accuracy}
\begin{center}
\resizebox{\textwidth}{!}{%
\begin{tabular}{c c c c c c c c c c c c c c c c}
\hline
\textbf{Model} & \textbf{SENet} & \textbf{EfficientNet} & \textbf{ResNet-34} & \textbf{VGG11} & \textbf{ShuffleNet-v2} & \textbf{DenseNet121} & \textbf{BERT-tiny} & \textbf{BERT-tiny} & \textbf{BERT-base} & \textbf{BERT-base} & \textbf{BERT-base} & \textbf{BERT-base} & \textbf{BERT-base} & \textbf{BERT-base} \\
Dataset & Cifar10 & Cifar10 & ImageNet & ImageNet & ImageNet & ImageNet & stsb$^+$ & mnli & cola* & mrpc & rte & sst2 & qnli & qqp \\
\hline
Baseline (FP32)         & 95.4  & 91.17 & 73.3  & 70.38 & 69.35 & 74.43 & 0.806 & 69.9  & 0.585 & 84.55 & 64.98 & 92.43 & 91.54 & 90.9\\
\hline
FP32 + crossbar noise          & 95.4  & 91.14 & 73.17 & 70.21 & 68.9  & 74.15 & 0.807 & 69.93 & 0.585 & 84.55 & 65.7  & 92.43 & 91.46 & 90.9\\
\ballnumber{1}    & 95.4  & 91.17 & 73.3  & 70.22 & 69.01 & 74.15 & 0.81  & 69.92 & 0.586 & 84.73 & 65.79 & 92.6  & 91.98 & 90.99\\
\ballnumber{2}          & 94.67 & 90.81 & 73.15 & 69.96 & 68.54 & 72.39 & 0.805 & 69.86 & 0.597 & 83.57 & 65.34 & 92.2  & 89.4  & 88.89\\
\hline
\ballnumber{3}          & 93.86 & 89.52 & 72.92 & 69.35 & 68.22 & 72.42 & 0.81  & 68.13 & 0.57  & 82.35 & 64.25 & 92.31 & 89.09 & 89.12\\
\ballnumber{3} + ACAM noise          & 61.92 & 54.06 & 41.94 & 50.53 & 61.69 & 53.23 & 0.701 & 49.00 & 0.461 & 78.84 & 61.73 & 90.71 & 86.28 & 87.41\\
\ballnumber{4}    & 93.23 & 89.05 & 72.06 & 67.4  & 68.05 & 71.34 & 0.805 & 68.58 & 0.58  & 84.55 & 64.33 & 92.43 & 90.32 & 89.19\\
\hline
\end{tabular}
}
\begin{flushleft}
\scriptsize{$^+$ = Pearson Correlation Coefficient (PCC), range [-1, 1], higher is better. \quad\quad  * =  Matthews Correlation Coefficient (MCC), range [-1, 1], higher is better.}\\
\vspace{-0.2in}
\end{flushleft}
\end{center}
\end{table*}

We select a set of popular AI models for evaluation.
For the CNNs, we tested CIFAR-10~\cite{krizhevsky2009learning} (on SENet~\cite{hu2018squeeze}, and EfficientNet~\cite{tan2019efficientnet}), and ImageNet~\cite{deng2009imagenet} (on ResNet-34~\cite{he2016deep}, VGG11~\cite{simonyan2014very}, ShuffleNet-V2~\cite{ma2018shufflenet} and DenseNet-121~\cite{huang2017densely}).
For LLMs, we use the GLUE~\cite{wang2019glue} dataset and test on the popular BERT model of different configurations (BERT-tiny~\cite{jiao2019tinybert} and BERT-base~\cite{devlin2019bert}).
We also use LLaMA3.2~\cite{grattafiori2024llama}, a state-of-the-art LLM with billions of parameters, to test NL-DPE's scalability in a multi-chip setting.

\section{Results}
\subsection{Design Space Exploration and Trade-offs}
\begin{figure}[htbp]
\centerline{\includegraphics[width=\linewidth]{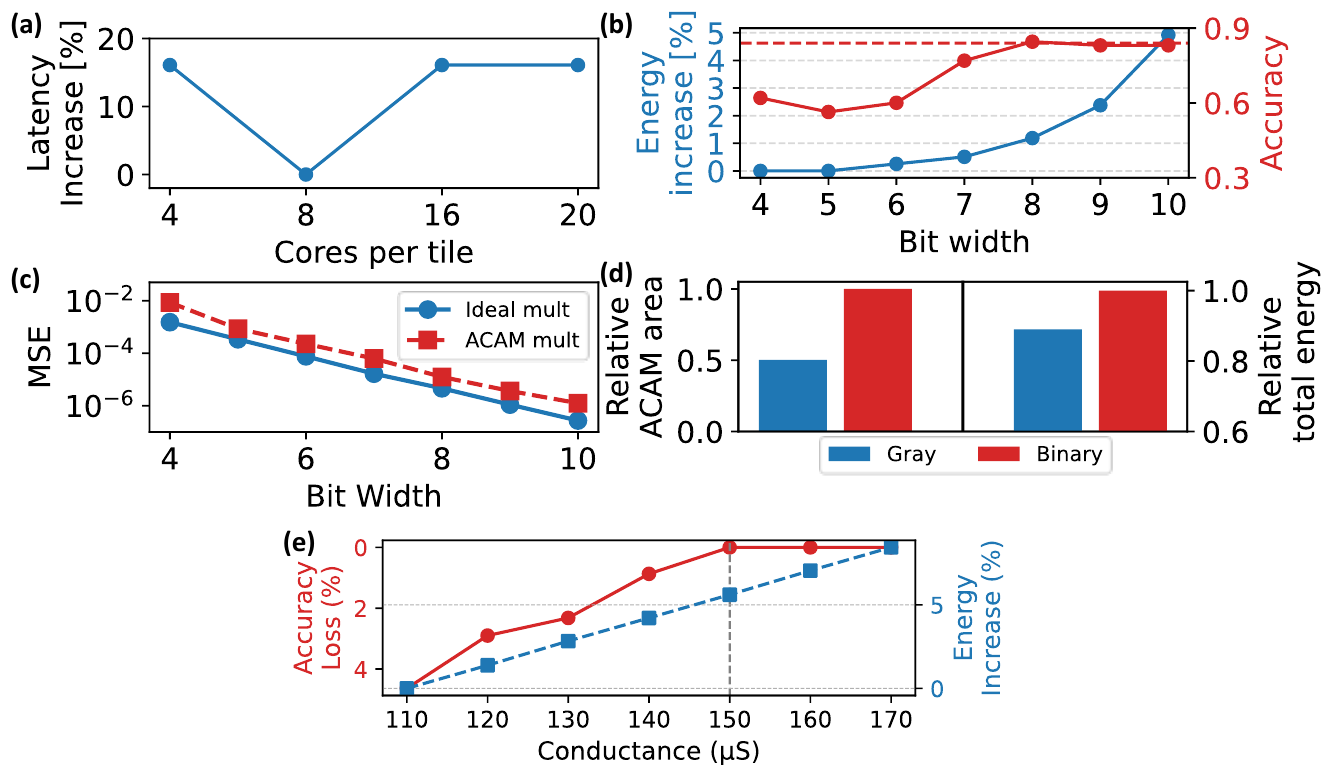}}
\caption{\textbf{(a)}Latency increase of NL-DPE accelerator with varying core counts per tile (normalized to the 8-core configuration).\textbf{(b)} Energy and accuracy as a function of bit width. \textbf{(c)} Mean Square Error (MSE) performing multiplication as a function of bit width for ideal software multiplier and ACAM-based multiplier.\textbf{(d)} Relative ACAM area and total energy for Gray and Binary encoding. textbf{(e)} Tradeoff between accuracy, energy consumption and RRAM's maximum conductance on BERT-base.}
\vspace{-0.1in}
\label{fig:dse}
\end{figure}
We begin by performing design space exploration to determine key parameters and investigate trade-offs in our accelerator architecture.
All results are based on BERT-base running the mrpc task, with other benchmarks showing similar trends.

\subsubsection{Cores in Tile} \label{sec:tradeoff-cores}

We evaluate inference latency for different core counts per NL-DPE tile (Fig.~\ref{fig:dse}(a), Table~\ref{tab:architecture}).
Latency is normalized to the 8-core setup, which achieves the best performance.
Fewer cores underutilize parallelism, while more cores face resource contention and off-chip memory access.
Thus, we adopt 8 cores per tile for all experiments.

\subsubsection{Bit width exploration} \label{sec:tradeoff-bitwidth}

We study the impact of bit width on NL-DPE.
Higher bit widths improve accuracy but increase ACAM usage and energy.
Below 7 bits, accuracy drops $>50\%$ even with NAF.
At 8 bits, accuracy matches full precision with $<1\%$ energy increase than 7 bits, and higher widths offer no gain.
We adopt 8-bit precision for NL-DPE.

Fig.~\ref{fig:dse}(c) shows the MSE between ACAM DT-based multiplication and ideal software computation across bit widths.
An 8-bit ACAM multiplier achieves precision comparable to a 7-bit digital multiplier.

\subsubsection{Gray encoding} \label{sec:tradeoff-encoding}

We evaluate the impact of Gray encoding on area and energy, as shown in Fig.~\ref{fig:dse}(d).
Gray encoding reduces the ACAM size by 50\% at the cost of a few additional XOR gates, whose area and energy overheads are negligible compared to the savings from the smaller ACAM, as shown by the relative energy savings.
We evaluated the inference accuracy for binary and Gray encoding, without seeing significant differences (e.g., both reach software equivalent accuracy in the case of BERT-base after NAF.)

\subsubsection{Conductance range} \label{sec:tradeoff-conductance}
Fig.~\ref{fig:dse}(e) shows the trade-off between RRAM conductance range, accuracy, and energy.
Increasing the maximum conductance improves precision and reduces noise impact, but raises energy.
Accuracy saturates beyond 150$\mu$S, so we adopt a 0.01–150$\mu$S range for all crossbar and ACAM RRAMs.

{\ccblue
\subsection{ACAM vs. Digital Function Units}
}

We evaluate ACAM as a replacement for ADC and activation units, comparing it to an 8-bit ADC~\cite{wang2025single} and FlexSFU~\cite{reggiani2023flex} at 16nm.
Table~\ref{tbl:acam-adc-sfu} shows 98\% lower power and $86\times$ smaller area at a constant latency, enabling each crossbar column to have a dedicated ACAM for maximum performance.

\begin{table}[ht]
\centering
\caption{ACAM vs. ADC+Activation.}
\scriptsize 
\label{tbl:acam-adc-sfu}
\begin{tabular}
{|c!{\vrule width 2pt}c|c|c!{\vrule width 2pt}c!{\vrule width 2pt}c|}
\hline
& \textbf{ADC} & \textbf{Activations} & \textbf{Total} & \textbf{ACAM} & \textbf{Improvement}\\
\hline
\textbf{Power [$mW$]} & 5.73 & 3.38 & 9.11 & 0.17 & $\sim$98\% better \\
\hline
\textbf{Area [$\mu m^2$]} & 457 & 20867 & 21324 & 247 & $\sim\times$86 better \\
\hline
\end{tabular}
\end{table}

\begin{table}[ht]
\centering
\caption{ACAM vs. Digital Multipliers.}
\scriptsize 
\label{tbl:rev1-acam-mul}
\begin{tabular}
{|c!{\vrule width 2pt}c|c|c|}
\hline
& \textbf{Fixed Point} & \textbf{Floating Point} & \textbf{ACAM}\\
\hline
\textbf{Power [$\mu W$]} & 539 & 1582  & 513 \\
\hline
\textbf{Area [$\mu m^2$]} & 1059 & 43610  & 774 \\
\hline
\end{tabular}
\vspace{-0.2in}
\end{table}

{\ccblue
Table~\ref{tbl:rev1-acam-mul} compares our 8-bit ACAM multiplier with conventional 8-bit digital designs, all scaled to 16nm.
Compared to the fixed-point design~\cite{mondal2023approximate}, it reduces power by 5\% and area by 27\%.
Unlike digital multipliers, ACAM directly processes continuous analog crossbar outputs without ADCs, supports flexible output formats (e.g., Float8, BFloat8), so we also compare with a floating-point design~\cite{computingfpnew} and achieves 68\% lower power and 98\% smaller area.
}

{\ccblue
\subsection{Ramp vs Flash non-linear ADC}
Table~\ref{tab:ADC_1} reports the comparison  with the 16nm ramp ADC presented in \cite{yang2025efficient}, which is also capable of computing the activation functions while performing the conversion. 
When comparing the conversion speed and latency of the two ADCs, the flash topology we propose has a clear edge: the conversion happens in a single search cycle, and the latency does not worsen exponentially with the number of bits. The circuit proposed in \cite{yang2025efficient} shows superior energy-per-conversion, for the 5-bit case; when normalizing by the number of levels of the conversion (Walden FOM, typically reported in ADC literature), the ramp ADC is 20-24\% more energy efficient. While the area of the ADC we propose is $\times$2.26 smaller, for both topologies each DPE column can be paired to one ADC; the area overhead of the ramp ADC is mainly due to the analog circuit peripherals needed: the RRAM array accounts for $\sim$5\% of the area, with the op-amps and the integration capacitances taking most of the ramp ADC area.
}

{
\setlength{\tabcolsep}{3pt}
\begin{table}[h]
\centering
\caption{Ramp vs Flash non-linear ADC}
\scriptsize 
\label{tab:ADC_1}
\begin{tabular}{|cccccc|}
\hline
 &
  \textbf{ENOB} &
  \textbf{\begin{tabular}[c]{@{}c@{}}Sampl\\ Freq. [MHz]\end{tabular}} &
  \textbf{\begin{tabular}[c]{@{}c@{}}Energy per\\ conv. [pJ]\end{tabular}} &
  \textbf{\begin{tabular}[c]{@{}c@{}}Area per\\ Col. [um\textsuperscript{2}]\end{tabular}} &
  \textbf{\begin{tabular}[c]{@{}c@{}}Walden FOM\\ {[}fJ/levels{]}\end{tabular}} \\ \hline
\begin{tabular}[c]{@{}c@{}}This\\ work\end{tabular} &
    8 & 1,000  & 5.73 & 247 & 22.4 \\ \hline
\begin{tabular}[c]{@{}c@{}}\cite{yang2025efficient}\\ (5-bit)\end{tabular} &
    5 & 31.2  & 0.595 & 558 & 18.6 \\ \hline
\begin{tabular}[c]{@{}c@{}}\cite{yang2025efficient}\\ (8-bit)\end{tabular} &
     8 & 3.9  & 4.6 & 558 & 18.0 \\ \hline
\end{tabular}%
\vspace{-0.2in}
\end{table}
}

\subsection{Accuracy}
Table~\ref{tbl:accuracy} shows the accuracy of various models at different stages of NAF, except for stsb and cola datasets, which use Pearson Correlation Coefficient (PPC) and Matthews Correlation Coefficient (MCC) as metrics. 
The ball numbers correspond to the steps depicted in Fig.~\ref{fig:finetune}.
Only a small error is introduced by the crossbars even in the presence of noise (FP32 + crossbar noise and \ballnumber{1}), thanks to the mitigation effect of A-SL.
Using DT-trained piecewise activations (\ballnumber{3}) doesn't introduce significant error as well, making the approach potentially generalizable to other accelerators that may efficiently perform inference of tree-based models~\cite{pedretti2021tree}.
However, in the presence of ACAM noise, the DTs experience significant accuracy degradation (as shown by \ballnumber{3} + ACAM noise), which can only be mitigated through ACAM NAF (\ballnumber{4}).
Thus, results demonstrate that our DT-trained activations approach is feasible and does not impact accuracy, but NAF is needed for practical utilization with analog accelerators.

\subsection{Performance and Energy}
\begin{figure*}[tbp]
\begin{center}
\includegraphics[width=\linewidth]{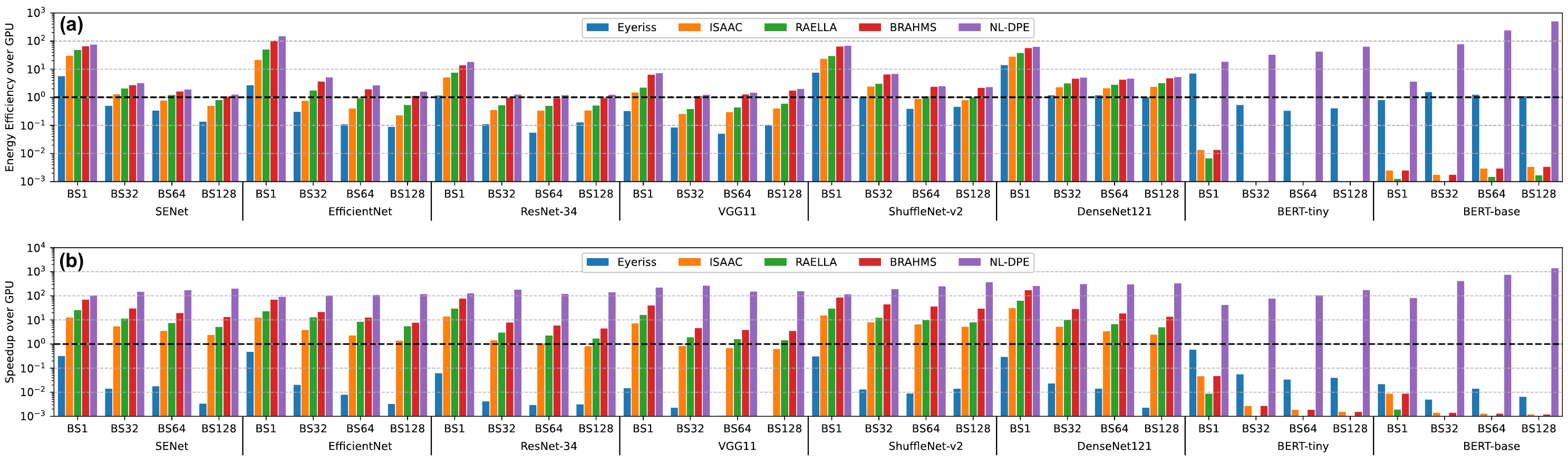}
\caption{(a) Normalized energy efficiency and (b) speedup compared to the GPU baseline}
\vspace{-0.3in}
\label{fig:speedup_energy}
\end{center}
\end{figure*}

Fig.~\ref{fig:speedup_energy} shows NL-DPE speedup and energy efficiency versus GPU across batch sizes.
For single-sample inference ($BS=1$), GPU suffers from low utilization, while NL-DPE is 112$\times$ faster and 28$\times$ more energy-efficient. Compared to other IMC accelerators, NL-DPE outperforms by replacing bulky ADCs/VFUs and natively supporting attention, reducing digital post-processing latency.

We evaluate increasing batch sizes.
While GPUs benefit from large batches, conventional IMC accelerators see limited gains due to shared ADCs and VFUs, causing high latency and bottlenecks.
In contrast, NL-DPE maintains high speedup and energy efficiency across all batches: each crossbar column has a dedicated ACAM for ADC conversion, reducing conversion latency to 3ns, and ACAM units compute multiple DMMul and activations in parallel.
Eyeriss suffers from repeated memory accesses.
NL-DPE consistently outperforms all baselines on BERT, achieving 249$\times$ speedup over GPU for multi-batch inference.

\begin{figure}[htbp]
\vspace{-0.1in}
\centerline{\includegraphics[width=0.7\linewidth]{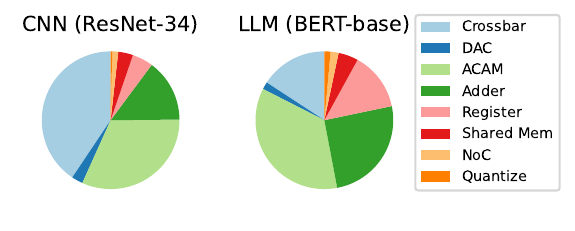}}
\vspace{-0.1in}
\caption{Energy breakdown for various components on a NL-DPE chip for representative CNN and LLM workloads.}
\vspace{-0.1in}
\label{fig:energy_breakdown}
\end{figure}

Fig.~\ref{fig:energy_breakdown} presents the energy breakdown of on-chip components in NL-DPE for a representative CNN workload (ResNet34) and LLM workload (BERT).
Most of the energy is consumed by the compute units (i.e., crossbars, ACAMs, and adders), highlighting the high computational intensity of the NL-DPE architecture.

Importantly, for LLM, NL-DPE is the only architecture improving the performance compared with modern GPUs, particularly for large batch sizes.
This is due to the unique capabilities of performing the attention mechanism by a series of highly efficient function approximations with ACAMs, without the need for rewriting the crossbar arrays or using VFUs.
Thus, the NL-DPE architecture distinguishes itself as the only IMC solution capable of scalable performance for contemporary, large-scale workloads. 

\subsection{Scalability for Larger LLMs} \label{sec:scalability}

\begin{figure}[htbp]
\vspace{-0.1in}
\centerline{\includegraphics[width=0.9\linewidth]{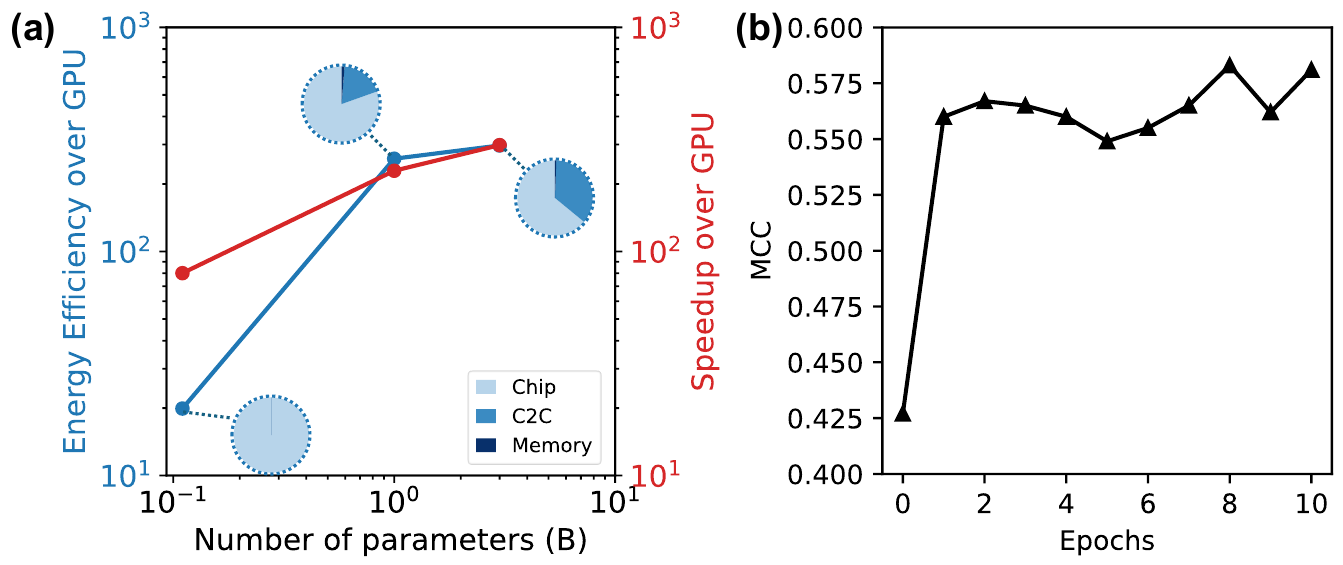}}
\caption{\textbf{(a)} Energy efficiency and speedup over GPU when scaling up NL-DPE for larger LLMs (i.e. LLaMA models). \textbf{(b)} MCC of cola on BERT-base as a function of NAF epochs.}
\vspace{-0.1in}
\label{fig:scalability}
\end{figure}

\subsubsection{Hardware Scalability} \label{sec:scale_efficiency}

As model size grows, a single chip cannot hold the entire model.
We adopt a multi-chip setup to avoid RRAM write endurance issues, rather than time-multiplexing, which incurs frequent writes.

BERT-base fits on one chip, so we evaluate scalability with LLaMA models: the 1B-parameter variant (LLaMA3.2-1B) uses 3 chips, and the 3B-parameter variant (LLaMA3.2-3B) uses 8 chips.
Chips are connected via a 10 Gbps chip-to-chip (C2C) interface with 30 pJ/bit energy cost, representing a conservative, worst-case scenario.

Even with this overestimated C2C overhead, NL-DPE achieves approximately 100$\times$ speedup and energy efficiency improvement over GPUs (Fig.~\ref{fig:scalability}(a)).
Pie charts show C2C energy rises with chip count (18\% for 1B, 35\% for 3B), while most energy remains within NL-DPE chips.
Off-chip memory contributes negligibly ($<1.5\%$) to total energy consumption due to IMC’s weight-stationary design.

\subsubsection{NAF Scalability} \label{sec:training_overhead}

Scalability is a key concern for in-memory accelerators, especially regarding in-device or end-to-end finetuning.
We next evaluate the overhead of our proposed NAF for larger models.

Our results show NL-DPE with NAF is highly scalable.
NAF is software-based and applied before deployment, with each DT trained and fine-tuned independently, avoiding full LLM finetuning.
To approximate an activation function, each DT uses only 5000 randomly sampled inputs, completing training in seconds via scikit-learn.
Fine-tuning uses a small input set for a few epochs; 10 epochs suffice to recover accuracy lost to ACAM noise (Fig.~\ref{fig:scalability}(b)).
Since DTs are lightweight and independent, multiple DTs can be processed in parallel, ensuring high scalability.

{\ccblue
\subsection{Noise Analysis}

\subsubsection{ACAM-Based Functions}

We assess ACAM-based computation by plotting the correlation between its outputs and ideal digital results.

\begin{figure}[htbp]
\centerline{\includegraphics[width=\linewidth]{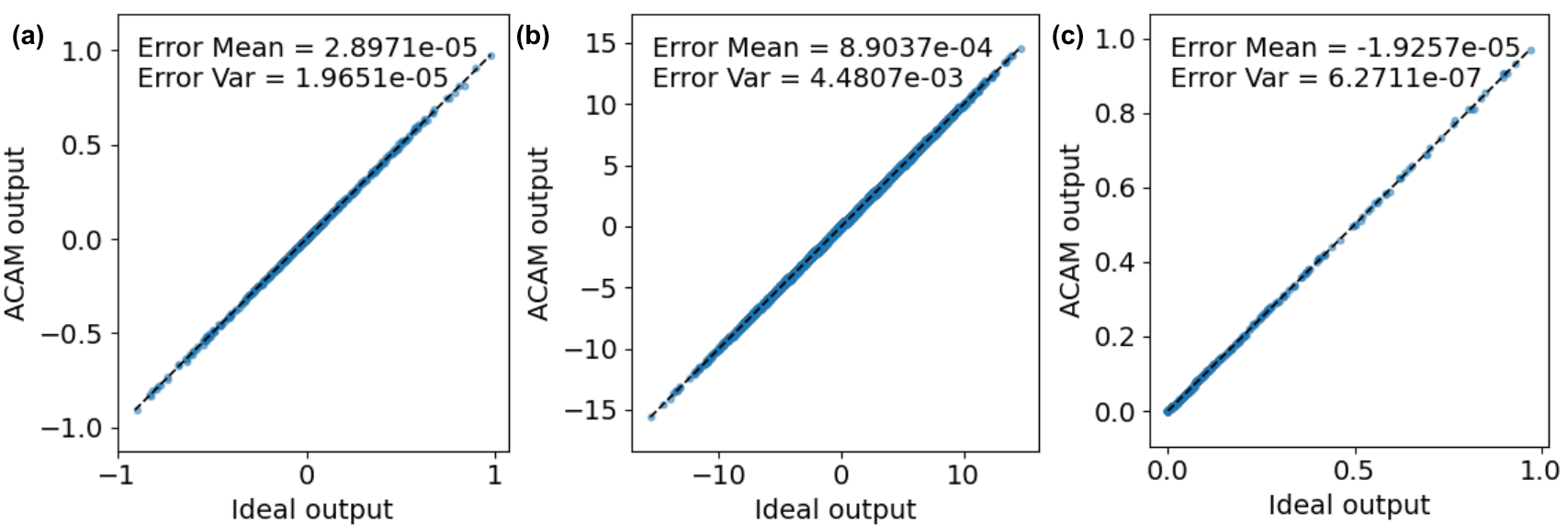}}
\caption{Correlation between the outputs of the proposed ACAM-based units and the ideal outputs: \textbf{(a)} multiplication results, \textbf{(b)} matrix multiplication of two 256×256 matrices, \textbf{(c)} softmax outputs.}
\label{fig:rev1-acam_unit_error_dist}
\vspace{-0.2in}
\end{figure}

\begin{itemize}[leftmargin=*]
\item
Figure~\ref{fig:rev1-acam_unit_error_dist}(a) shows a strong linear correlation between the 8-bit ACAM multiplier and a conventional 8-bit digital multiplier across 500 inputs, with MSE $2.8971 \times 10^{-5}$ and variance $1.9651 \times 10^{-5}$.

\item 
Figure~\ref{fig:rev1-acam_unit_error_dist}(b) shows strong correlation between $256 \times 256$ ACAM-based matrix multiplication and the digital baseline, with MSE $8.9037 \times 10^{-4}$ and variance $4.4807 \times 10^{-3}$, reflecting realistic LLM workloads.

\item 
Figure~\ref{fig:rev1-acam_unit_error_dist}(c) further shows the softmax unit’s correlation with the digital baseline, exhibiting a mean error of $-1.9257 \times 10^{-5}$ and variance $6.2711 \times 10^{-7}$, confirming high numerical fidelity for complex computations.
\end{itemize}

These results demonstrate that our proposed ACAM-based computation, after NAF, achieves a precision very close to its digital counterpart.

\subsubsection{Variation of Noise Distribution}

To test NAF’s robustness to varying noise, we varied the noise standard deviation during inference.
Figure~\ref{fig:rev4-error_var_std} plots accuracy loss versus normalized noise standard deviation (0.5$\times$–2.5$\times$, baseline = 1.0).

\begin{figure}[htbp]
\vspace{-0.1in}
\centerline{\includegraphics[width=0.6\linewidth]{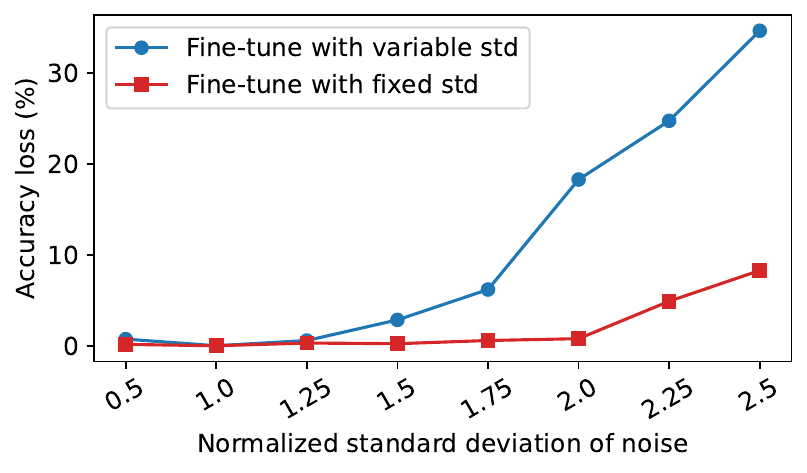}}
\caption{Accuracy loss of BERT-base on sst2 under different inference-time noise standard deviations. Averaged across 5 runs.}
\vspace{-0.1in}
\label{fig:rev4-error_var_std}
\end{figure}


\begin{itemize}[leftmargin=*]

\item
Scaled-noise training (blue curve): In this case, the training was performed with scaled noise std values corresponding to the inference-time settings.
We found that when the noise std is enlarged (e.g., $>$1.5$\times$), the model experiences noticeable accuracy degradation due to the reduced convergence stability during fine-tuning.

\item
Fixed-noise training (red curve): During fine-tuning, the model was trained with the hardware-characterized noise model (normalized std = 1.0) as proposed in our noise model, and only the inference-time noise was varied.
We observe that the accuracy remains stable even when the inference noise std is increased by up to 2$\times$, demonstrating that our noise-aware fine-tuning method generalizes well to different noise levels.
\end{itemize}

\noindent
These results suggest that our hardware-calibrated noise model not only eliminates the need for on-device fine-tuning but also provides sufficient generalization across different device noise levels, without requiring Gaussian noise injection during training.

\subsubsection{Stuck-At Faults}

Figure~\ref{fig:rev3-error_stuck} shows the impact of stuck-at faults (SAFs) on the crossbar and ACAM, and the tolerance of our Noise-Aware Fine-tuning (NAF), using BERT-base on SST-2 from the GLUE dataset.

\begin{figure}[htbp]
\vspace{-0.15in}
\centerline{\includegraphics[width=0.6\linewidth]{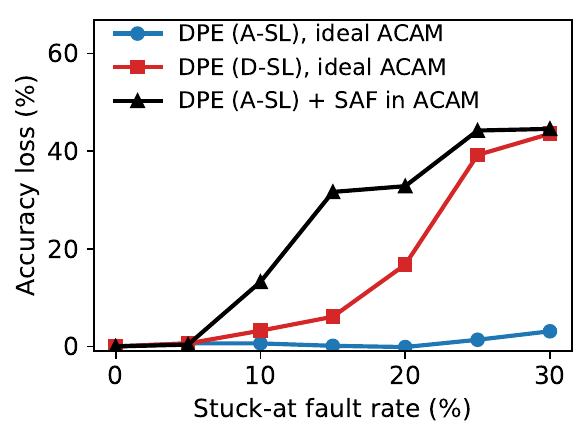}}
\vspace{-0.1in}
\caption{Accuracy degradation under stuck-at faults (SAFs) in crossbar and ACAM averaged across 5 runs. Red: crossbar SAFs with digital slicing. Blue: crossbar SAFs with analog slicing. Black: SAFs in both ACAM and crossbar (analog slicing).}
\vspace{-0.1in}
\label{fig:rev3-error_stuck}
\end{figure}

We first evaluate two crossbar weight mappings under 5–30\% SAFs, assuming ideal ACAM: conventional digital mapping (red) and analog slicing (blue).
Both maintain near-ideal accuracy up to 5\% faults, while analog slicing tolerates up to 20\% SAFs by distributing each weight across two RRAM cells, allowing fault-free cells to partially compensate for stuck cells.

Then we test SAF on ACAM, which shows higher sensitivity because (1) analog slicing is not applicable to its AND operations, and (2) each output bit is computed independently, so errors in higher bits cause exponentially larger deviations.

During NAF, two strategies are used to mitigate stuck-at faults (SAFs) on ACAM: (1) if a DT does not use all ACAM rows, data mapped to a faulty cell can be reassigned to an unused row, disabling the defective one by skipping its pre-charge during inference; (2) knowing SAF locations in advance, NAF avoids updating or assigning values to these cells during fine-tuning.
Figure~\ref{fig:rev3-error_stuck} (black) shows accuracy degradation under these optimizations, tolerating up to 5\% SAFs.
Additional methods, such as error correction coding (ECC)~\cite{roth2024access}, could further mitigate SAFs, left for future work.
}

\section{Discussion}~\label{sec:discussion}
{\underline{\bf NAF for multiple chips.}}
Unlike prior work requiring post-deployment device-in-the-loop fine-tuning, our NAF integrates a per-device, pre-deployment noise model by reading/writing different resistance values.
This eliminates fine-tuning on hardware, improves endurance, and minimizes overhead.
Noise extraction can be incorporated into fabrication, and software-based fine-tuning requires fewer than 10 epochs (Fig.~\ref{fig:scalability}(b)), keeping multi-chip calibration overhead minimal.

{\underline{\bf Conductance drift.}}
Our experiments show TaOx RRAMs have negligible conductance drift, with only stochastic fluctuations, making NAF highly effective.
For other memory types with drift, NAF can incorporate a drift model without changing the training algorithm (Algorithm~\ref{algo:naf}).

{\underline{\bf Endurance and wear-out analysis.}}
NL-DPE targets inference, with fine-tuned weights programmed once and reused.
Device programming during noise profiling requires only $\sim10s$ cycles~\cite{merced2016repeatable}, negligible compared to RRAM endurance of up to $10^8$ cycles~\cite{mannocci2023memory}.

{\underline{\bf Potential paths for improving the accuracy.}}
NAF fine-tunes each DT individually using synthetic data, keeping overhead low and avoiding original training data.
If original training data is available, end-to-end fine-tuning can further improve accuracy at higher cost.
DT ensembles (e.g., Random Forests, XGBoost) or error correction codes~\cite{roth2024access} can also enhance robustness against ACAM errors.

{\underline{\bf Other operators.}}
Few non-critical operations may need special handling.
In LLMs, layer normalization can run on a VFU, though it can be replaced by Dynamic Tanh~\cite{zhu2025transformers}, which ACAM handles efficiently.
In CNNs, max pooling is the only exception, using simple comparators.

{\underline{\bf Sparse models.}}
Sparsity is orthogonal and can be applied on top of our design. For example, SRE~\cite{yang2019sparse} improves efficiency by activating crossbars at finer granularity, reducing ADC precision needs. Similarly, when applied to NL-DPE, this can reduce ACAM usage.

\section{Related work}~\label{sec:related-work}
{\underline{\bf DPE-ACAM hybrid architectures.}}
BRAHMS~\cite{song2021brahms} uses ACAM to replace ADCs and project analog signals into non-linear outputs, first routing inputs through pass-gates to produce ReLU-like activations.
A/D conversion and projection are separate, adding overhead (Fig.~\ref{fig:speedup_energy}).
Fuse-and-Mix~\cite{zhu2022fuse} merges projection and A/D conversion, using ACAM as an address decoder for a memory lookup.
In contrast, NL-DPE outputs the result directly from ACAM, reducing peripherals, and supports arbitrary activations beyond ReLU$-\alpha$.

{\underline{\bf In-memory computing of the attention mechanism.}}
SPRINT~\cite{yazdanbakhsh2022sparse} uses analog comparators to prune low attention scores in crossbars, avoiding ADCs.
ReTransformer~\cite{yang2020retransformer} reduces crossbar writes when computing attention by avoiding intermediate storage (e.g., $\mathbf{K^T}$).
Both approaches are challenging on current RRAM due to limited endurance ($<10^9$ cycles)~\cite{mannocci2023memory}.
ReTransformer also approximates Softmax via logarithms and maps exp/log with logic-in-memory, but this requires crossbar writes per iteration, limiting practical deployment~\cite{shirinzadeh2017endurance}.
Kernel approximation~\cite{buchel2024kernel} reduces computation but adds overhead for element-wise multiplications, exponentials, and L2-norm in FP32.

\vspace{-0.1in}
\section{Conclusion}
We present NL-DPE, an ADC-less analog IMC primitive that unifies crossbars for VMMs and ACAMs for non-VMMs, enabling major CNN and LLM computations entirely in the analog domain.
Non-VMMs are approximated with DTs and efficiently processed by ACAMs.
A software-based NAF improves accuracy under RRAM noise.
Compared to conventional DPEs and GPUs, NL-DPE achieves 28$\times$ energy efficiency and 249$\times$ speedup on average.
NL-DPE offers a new building block for in-memory computing, fully leveraging programmable analog non-linear operations.
\clearpage

\bibliographystyle{IEEEtran}
\bibliography{refs}
\vspace{-0.2in}

\begin{IEEEbiography}
[{\vspace{-0.3in}\includegraphics[width=1in,height=1.25in,clip,keepaspectratio]{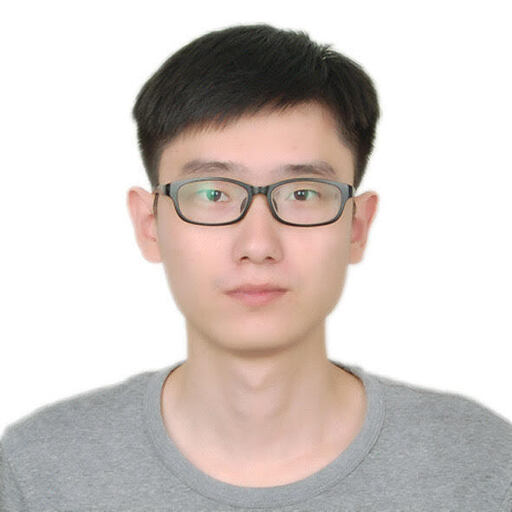}}]{Lei Zhao} is a Postdoctoral Researcher at Hewlett Packard Labs. He earned his Ph.D. in Computer Science from the University of Pittsburgh in 2022. His research centers on designing efficient neural network accelerators, with work spanning hardware architecture, compiler technologies, and system-level co-design for machine learning.
\end{IEEEbiography}
\vspace{-0.5in}

\begin{IEEEbiography}
[{\vspace{-0.3in}\includegraphics[width=1in,height=1.25in,clip,keepaspectratio]{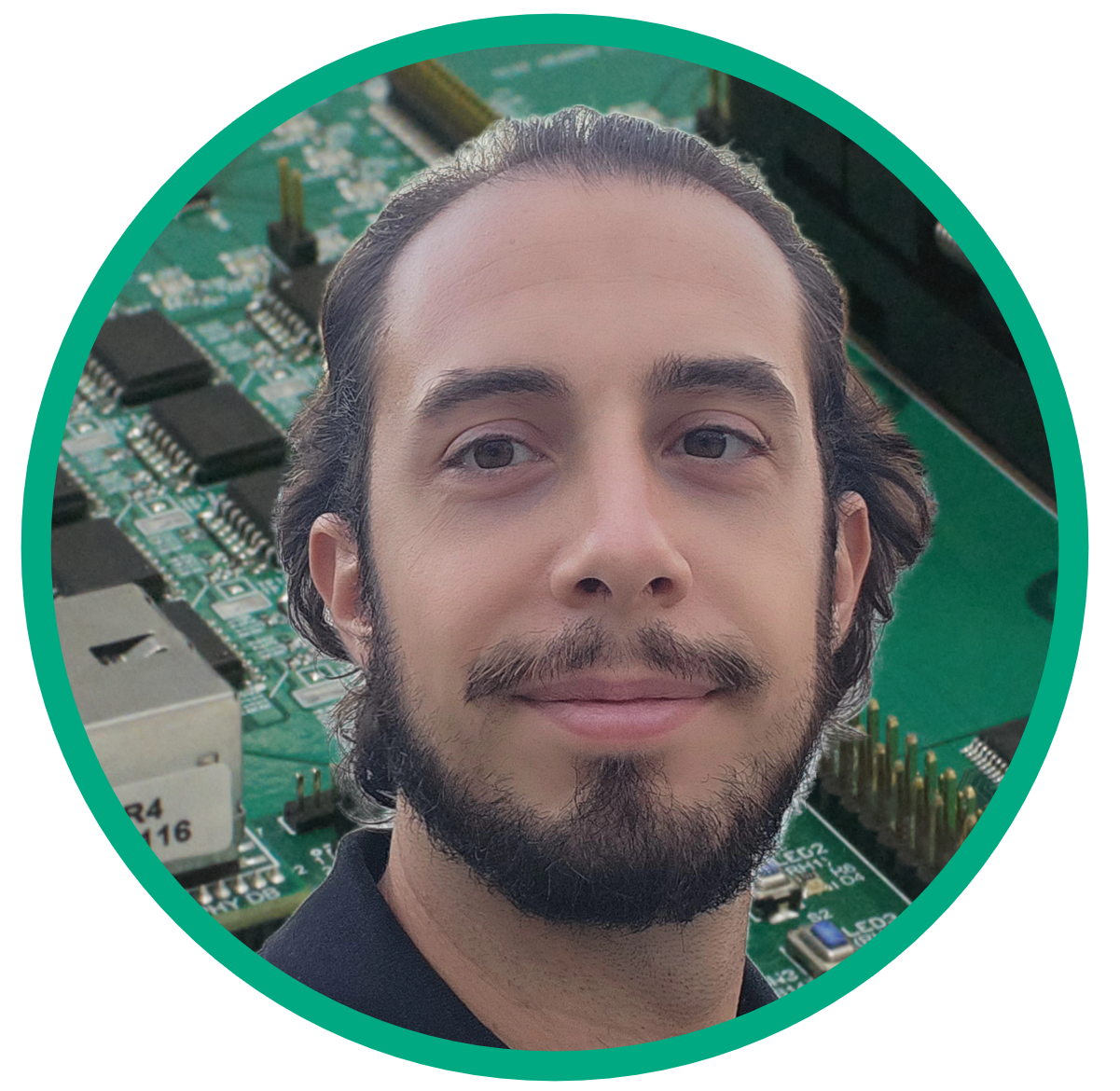}}]{Luca Buonanno} leads projects on AI-accelerator chiplets, in-memory computing for transformers, AI for networks, and optical computing. He earned a PhD summa cum laude in Information Technology from Politecnico di Milano in 2022 and joined Hewlett Packard Labs in 2021. His expertise spans VLSI design, nuclear instrumentation, and hardware accelerators, with over 50 publications and patents.
\end{IEEEbiography}
\vspace{-0.5in}

\begin{IEEEbiography}
[{\vspace{-0.3in}\includegraphics[width=1in,height=1.25in,clip,keepaspectratio]{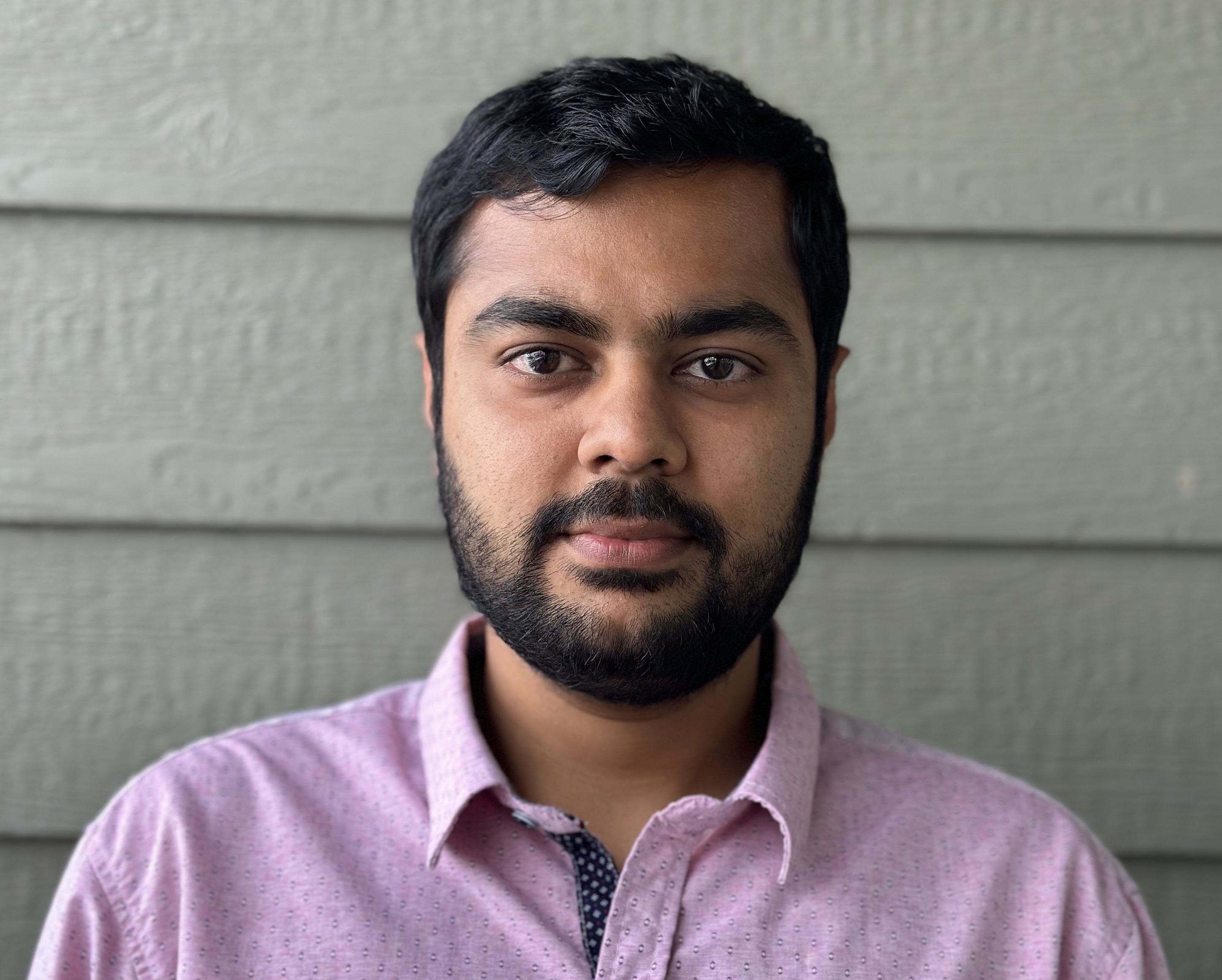}}]{Archit Gajjar} received his Ph.D. in Computer Engineering from North Carolina State University and is a postdoctoral research associate at Hewlett Packard Labs. He holds an M.S. from the University of Houston–Clear Lake and a B.Tech. from DA-IICT, India. His research focuses on applied AI/ML on customized hardware, FPGA accelerators, and high-level synthesis.
\end{IEEEbiography}
\vspace{-0.5in}

\begin{IEEEbiography}
[{\vspace{-0.3in}\includegraphics[width=1in,height=1.25in,clip,keepaspectratio]{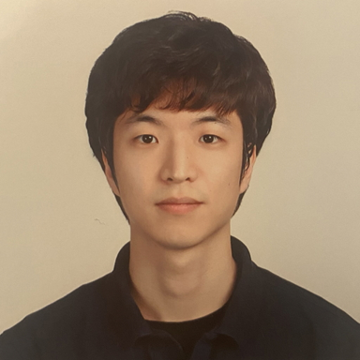}}]{John Moon} is a Research Scientist at Hewlett Packard Enterprise Labs. He received his Ph.D. degree in 2021 from University of Michigan, Ann Arbor, MI, USA. His research includes designing an in-memory computing accelerator, simulating architectures, and applying it to AI/ML.
\end{IEEEbiography}
\vspace{-0.5in}

\begin{IEEEbiography}
[{\vspace{-0.3in}\includegraphics[width=1in,height=1.25in,clip,keepaspectratio]{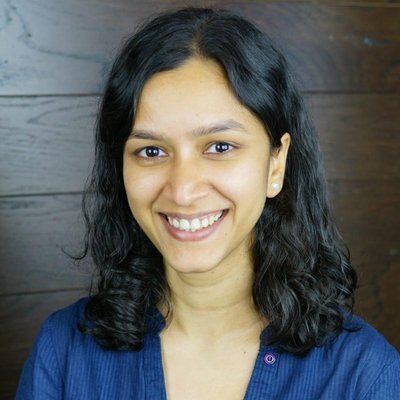}}]{Aishwarya Natarajan} is a Research Scientist at Hewlett Packard Labs, developing next-generation hardware accelerators using in-memory and neuromorphic computing. She received her Ph.D. in Electrical and Computer Engineering from Georgia Tech in 2021. Her research interests include emerging computing architectures and energy-efficient, analog circuits and systems..
\end{IEEEbiography}
\newpage

\begin{IEEEbiography}
[{\vspace{-0.3in}\includegraphics[width=1in,height=1.25in,clip,keepaspectratio]{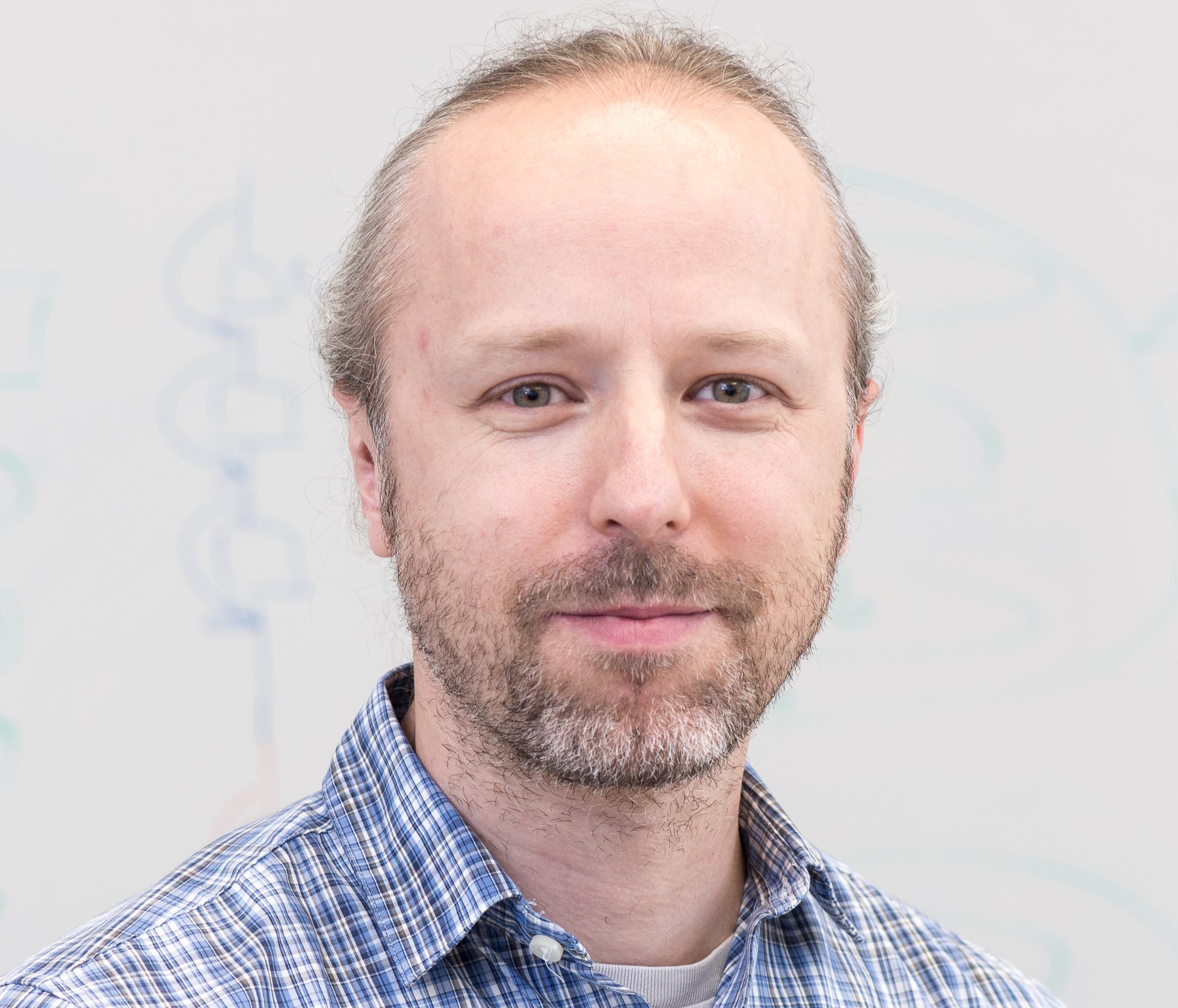}}]{Sergey Serebryako} is an AI/ML Engineer at Hewlett Packard Labs, focusing on NLP, deep learning benchmarking, and anomaly detection in telemetry from HPC and manufacturing systems. He represents HPE in MLCommons, co-chairing the MLCube Working Group, and applies time-series foundation models for forecasting and anomaly detection.
\end{IEEEbiography}
\vspace{-0.5in}

\begin{IEEEbiography}
[{\vspace{-0.2in}\includegraphics[width=1in,height=1.25in,clip,keepaspectratio]{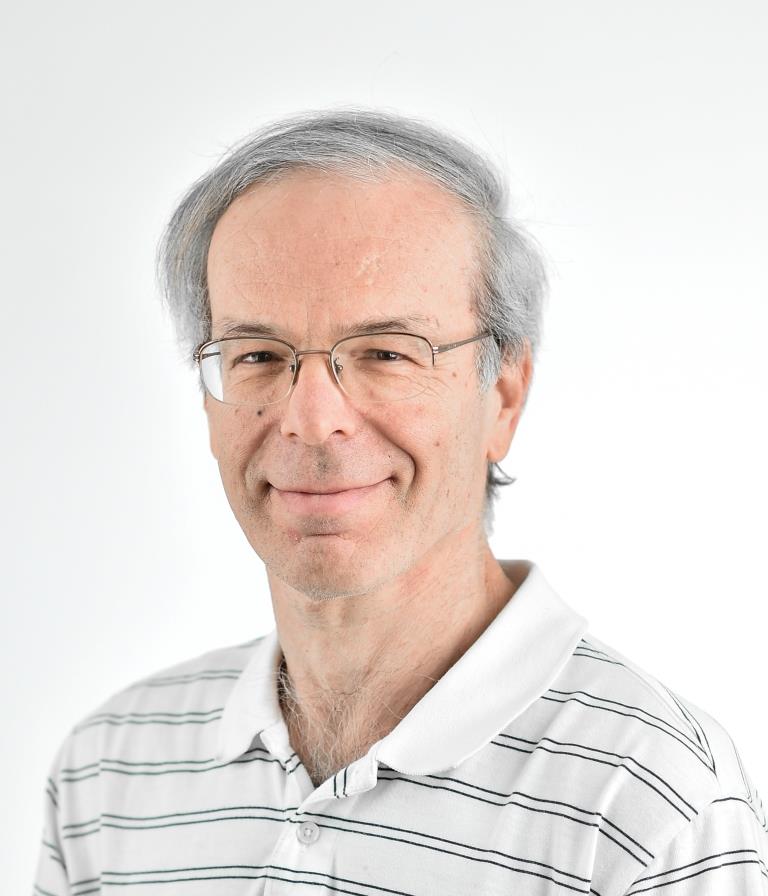}}]{Ron M. Roth} (Life Fellow, IEEE) received his B.Sc., M.Sc., and D.Sc. degrees from the Technion in Haifa, Israel, where he has been a faculty member in the Computer Science Department since 1988 and holds the General Yaakov Dori Chair in Engineering. He is the recipient of the 2021 NVMW Persistent Impact Prize for ECC. His research focuses on coding theory, information theory, and applications in storage, computation, and complexity.
\end{IEEEbiography}
\vspace{-0.5in}

\begin{IEEEbiography}
[{\vspace{-0.3in}\includegraphics[width=1in,height=1.25in,clip,keepaspectratio]{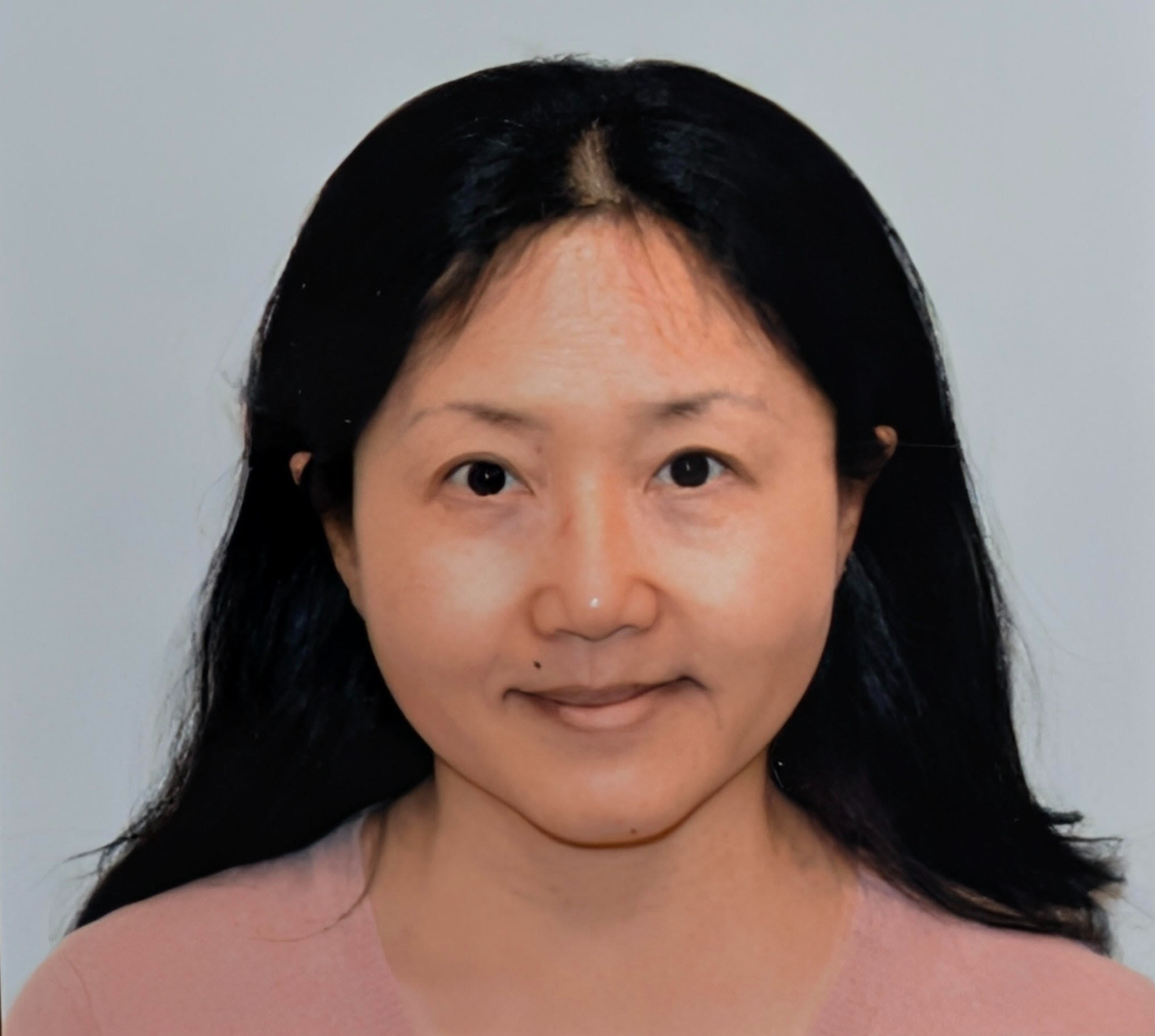}}]{Xia Sheng} received her B.E. and M.E. from Nankai University and her Ph.D. from Tokyo University of Agriculture and Technology. She joined Hewlett Packard Labs in 2000, focusing on thin-film technology and semiconductor devices for data storage, displays, and accelerated computing. She received the JSAP Young Scientist Award (1998) and IDW Best Paper Award (2011).
\end{IEEEbiography}
\vspace{-0.5in}

\begin{IEEEbiography}
[{\vspace{-0.3in}\includegraphics[width=1in,height=1.25in,clip,keepaspectratio]{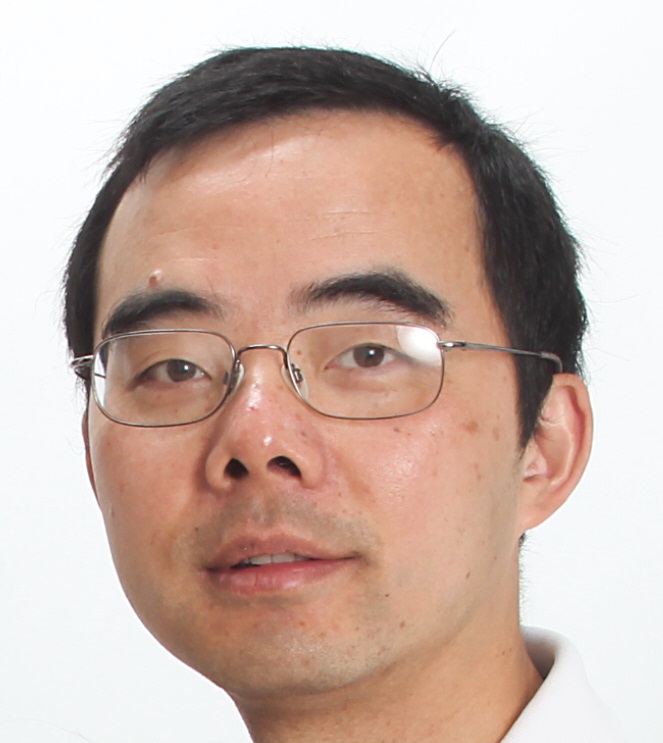}}]{Youtao Zhang} is a Professor of Computer Science at the University of Pittsburgh since 2006. He earned his Ph.D. from the University of Arizona in 2002 and was previously an assistant professor at the University of Texas at Dallas. His research covers computer security, program analysis, compiler optimization, and computer architecture. He is a member of ACM and IEEE.
\end{IEEEbiography}
\vspace{-0.5in}

\begin{IEEEbiography}
[{\vspace{-0.3in}\includegraphics[width=1in,height=1.25in,clip,keepaspectratio]{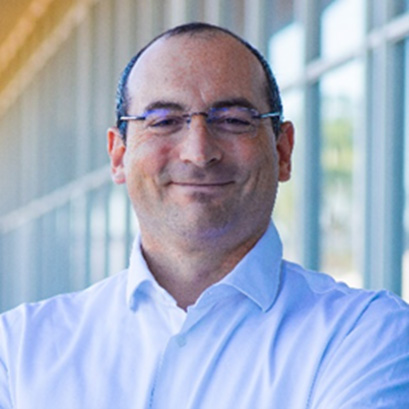}}]{Paolo Faraboschi} serves as VP in System Engineering at Arm. He was previously VP and Fellow at HPE, directing AI Research. He has worked on a broad range of technologies, from embedded VLIW processors, to low-power servers, and exascale supercomputers. He is an IEEE Fellow, with 70+ patents, and 100+ publications. He received a Ph.D. in EECS from the University of Genoa, Italy.
\end{IEEEbiography}
\vspace{-0.5in}

\begin{IEEEbiography}
[{\vspace{-0.3in}\includegraphics[width=1in,height=1.25in,clip,keepaspectratio]{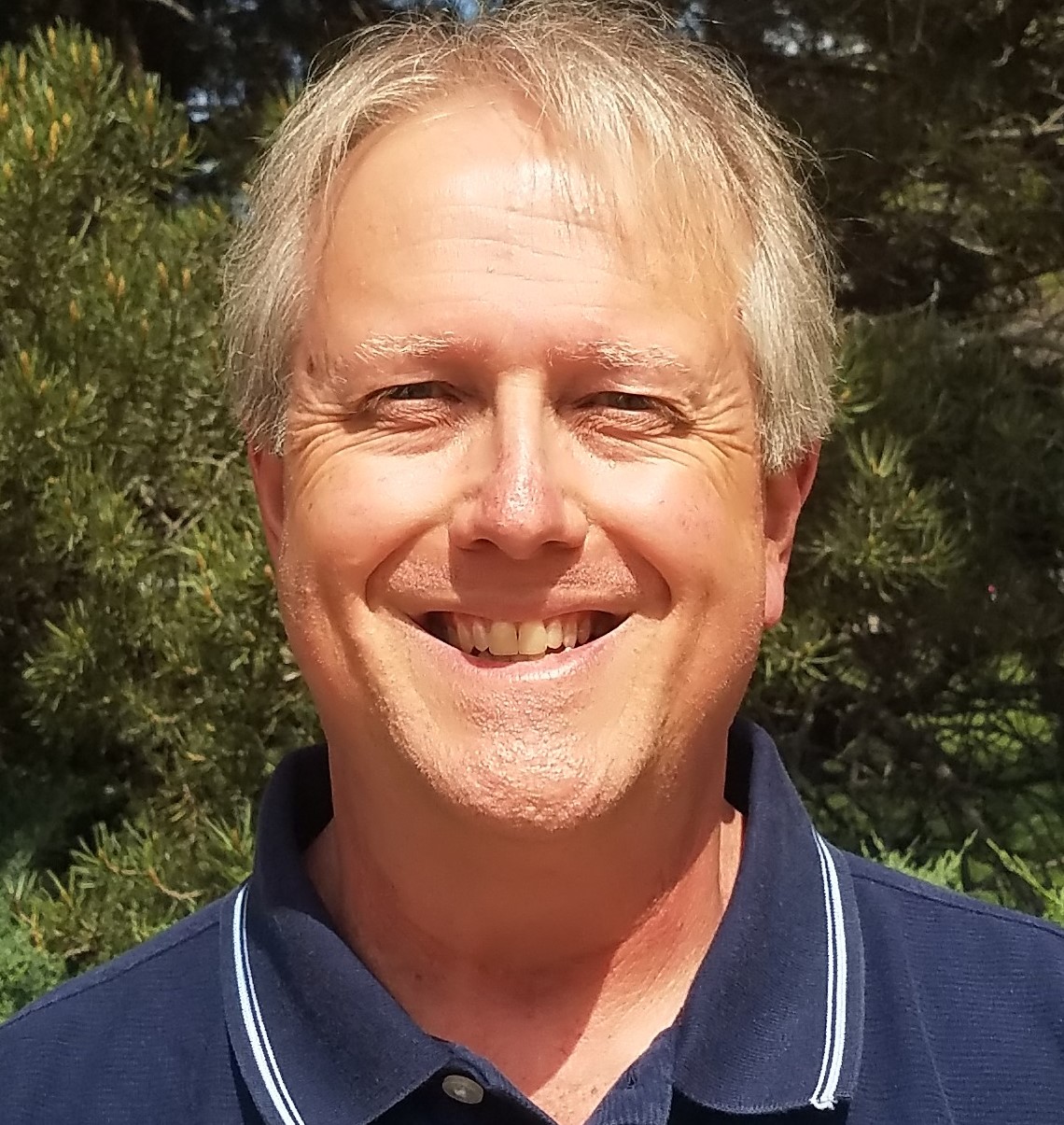}}]{Jim Ignowski} is the Director of Emerging Accelerators Research at HPE Labs. He has spent over 30 years with HP/HPE in R\&D across analog ICs, VLSI, and emerging non-volatile memory technologies. He holds an MSEE from Stanford University and a BSEE from Rice University.
\end{IEEEbiography}
\vspace{-0.5in}

\begin{IEEEbiography}
[{\vspace{-0.3in}\includegraphics[width=1in,height=1.25in,clip,keepaspectratio]{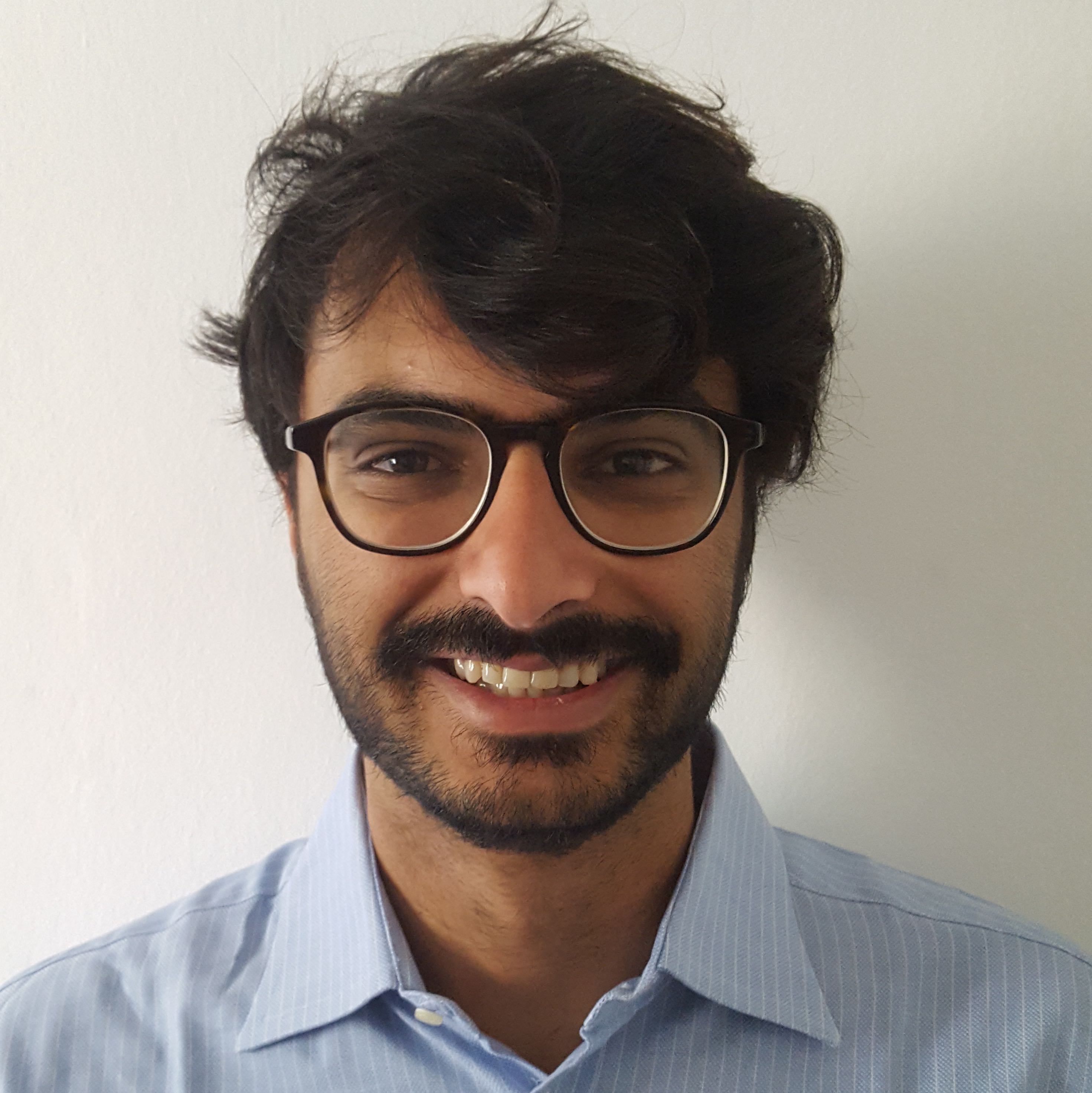}}]{Giacomo Pedretti} is a Principal AI/ML Researcher at Hewlett Packard Labs. His research interests include analog                   and physics-inspired hardware accelerators for AI and optimization. He is an IEEE and ACM member. He received his Ph.D. in Electrical Engineering from Politecnico di Milano in 2020.
\end{IEEEbiography}

\end{document}